\documentclass[aps,prd,superscriptaddress,nofootinbib,amsmath,amsfonts,groupedaddress]{revtex4-2}

\usepackage{setspace}
\usepackage{graphicx}
\usepackage{subfig}
\usepackage{epsfig}
\usepackage{bm}
\usepackage{amssymb}
\usepackage{float}
\usepackage{amsmath}
\usepackage{dcolumn}
\usepackage[colorlinks]{hyperref}
\usepackage[usenames,dvipsnames]{color}
\usepackage{enumitem}
\usepackage[autostyle]{csquotes}

\hypersetup{ breaklinks=true, pdfstartview={FitH}, colorlinks=true, linkcolor=blue, citecolor=red, filecolor=magenta, urlcolor=blue, anchorcolor=green, linktocpage=true }

\renewcommand{\(}{\left(}

\def\Om{\Omega}

\def\doi{http://doi.org}

\newcommand{\be}{\begin{equation}}
\newcommand{\ee}{\end{equation}}
\newcommand{\beano}{\begin{eqnarray*}}
\newcommand{\eeano}{\end{eqnarray*}}
\newcommand{\ba}{\begin{eqnarray}}
\newcommand{\ea}{\end{eqnarray}}

\begin{document}

\title{ EDSFD parametrization in $ f(R,T) $ gravity with linear curvature terms}

\author{J. K. Singh}
\email{jksingh@nsut.ac.in}
\affiliation{Department of Mathematics, Netaji Subhas University of Technology, New Delhi-110 078, India}

\author{Shaily}
\email{Shaily.ma19@nsut.ac.in}
\affiliation{Department of Mathematics, Netaji Subhas University of Technology, New Delhi-110 078, India}

\author{Harshna Balhara}
\email{harshna.ma19@nsut.ac.in}
\affiliation{Department of Mathematics, Netaji Subhas University of Technology, New Delhi-110 078, India}

\author{Sushant G. Ghosh} 
\email{sghosh2@jmi.ac.in}
\affiliation{Centre for Theoretical Physics, Jamia Millia Islamia, New Delhi 110025, India}
\affiliation{Astrophysics Research Centre, School of Mathematics, Statistics and Computer Science, University of KwaZulu-Natal, Private Bag X54001, Durban 4000, South Africa}

\author{Sunil D. Maharaj}
\email{maharaj@ukzn.ac.za}
\affiliation{Astrophysics Research Centre, School of Mathematics, Statistics and Computer Science, University of KwaZulu-Natal, Private Bag 54001,
Durban 4000, South Africa}

\begin{abstract}
This paper investigates the flat Friedmann-Lema$\hat{\imath}$tre-Robertson-Walker (FLRW) cosmological model using a suitable parameterization represented as a differential equation concerning the energy density of the scalar field, $\rho_{\phi}$, in the context of $f(R,T)$ gravity theory. This parameterization is known as the Energy Density Scalar Field Differential Equation (EDSFD) parametrization. It results in a solution of the Hubble parameter containing four model parameters, namely, $\Omega_{m0},\Omega_{\phi 0}, H_0,$ and $\alpha$. To constrain the model parameters, $77$ data points from the Hubble dataset, $1048$ points from the Pantheon dataset, and $6$ data points from BAO are used. Using the constrained values, we analyze and compare our model with the standard $\Lambda$CDM model. The evolution of the physical parameters, which includes the deceleration parameter, density parameter, Equation of State (EoS) for Dark Energy, and $Om(z)$ diagnostic, are discussed.   
\end{abstract}
\maketitle

\noindent
PACS numbers: {04.50.-h, 98.80.-k.}\\

Keywords: {Modified theory, EDSFD parametrization, Observational constraints}.

\section{ Introduction}\label{Sec1}
It is now generally accepted that the expansion of the cosmos at redshifts lower than unity underwent a \enquote{dynamic phase change} from a decelerating to an accelerating state, a phenomenon confirmed by numerous independent analyses \cite{BICEP2:2014owc, Planck:2015fie, Planck:2018vyg}. The measurements of the luminosity distance versus redshift relation using supernovae (SNeIa) of type Ia, which were initially interpreted in the context of $\Lambda$CDM scenarios using either background or inhomogeneous luminosity distances, provide the most direct evidence for the universe's current accelerating stage \cite{SST:1998fmf, SupernovaSearchTeam:2004lze, Riess:2006fw, SCP:1998vns, Weinberg:2013agg}. However, independent theoretical and observational analyses indicate the existence of more generic models featuring a negative-pressure dark energy component as their main component \cite{Padmanabhan:2002ji,  Peebles:2002gy, Copeland:2006wr}.

The current cosmic picture strongly supports the following: the spatial geometry (curvature) is flat, and the dynamics are controlled by a component known as dark energy (DE), which makes up $ 3/4^{th} $ of the composition and $ 1/4^{th} $ of the matter component (baryons plus dark). The simplest and most appealing way to characterize this dark energy component is to use a positive cosmological constant $ \Lambda $. The cosmological constant $\Lambda $ has an equation of state (EoS), $ \omega = \frac{p}{\rho} = -1 $, where $ p $ and $ \rho $ are the effective thermodynamic pressure and energy density associated with $\Lambda$. The dynamical DE scenario is an alternate DE proposal. Numerous scalar field dark energy models have been discussed so far, including the quintessence \cite{Peebles:1987ek, Ratra:1987rm, Singh:2022eun, Singh:2022nfm}, K-essence \cite{Armendariz-Picon:2000ulo}, tachyon \cite{Sen:2002in}, phantom \cite{Setare:2007eq}, ghost condensate \cite{Arkani-Hamed:2003pdi, Piazza:2004df}, quintom \cite{Setare:2007qu}, braneworld models \cite{Setare:2006sv}, and Chaplygin gas models \cite{Kamenshchik:2001cp} etc.

Modified theories of gravity have received much attention lately in light of their potential contribution to the universe's accelerating expansion. As a result of its simplicity and lack of ghosts, the so-called $ f(R) $ gravity theory \cite{Bertolami:2007gv, Bertolami:2007vu, Bertolami:2008ab, Harko:2008qz, Sotiriou:2008rp, Clifton:2011jh} is the most straightforward and most widely accepted classical extension of General Relativity (GR). In addition, compared to previous modifications of gravity, the reconstruction of $ f(R) $ theories that replicate the dark energy epoch and even the inflationary phase is typically simple \cite{DeFelice:2010aj}. It is possible for $ f(R) $ gravity to replicate the $ \Lambda $CDM era, imitate a cosmological constant in the present, and even unify all of cosmological history. Some $ f(R) $ gravities, or so-called viable models, can replicate a plausible cosmic evolution and pass the local gravitational criteria \cite{Nojiri:2010wj}. Studying $f(R)$ gravity models allows us to explore the connection between early-time inflation and late-time acceleration \cite{Nojiri:2008nt, Appleby:2009uf}. Existing literature highlights the crucial role played by higher-order curvature terms in $f(R)$ gravity models in preventing cosmological singularities \cite{Kanti:1998jd, Nojiri2008, Bamba:2008ut}.

In their work \cite{Nojiri:2003ni}, Nojiri and Odinstov examined a modified gravity theory that incorporates terms proportional to $ln(R)$ or $R^{-n}(ln R)^m$, which exhibit growth at low curvature. The $ln(R)$ or $R^{-n}(ln R)^m$ terms in $f(R)$ gravity models are known to contribute to the acceleration of the universe. Nojiri and Odinstov further investigated the cosmology of $f(R)$ gravity, specifically considering the form $f(R) = R+\gamma R^{-n}ln(\frac{R}{\mu^2})^m$ \cite{Nojiri:2006ri}. These functional forms were also utilized in previous studies \cite{Nojiri:2003ni, Nojiri:2006ri, Nozari:2009tv} to explore various aspects of the theory. Additionally, in \cite{Paul:2009nb}, it was demonstrated that these models exhibit the current accelerating phase of the universe, with the duration of this phase relying on the coupling constants of the gravitational action. The models of the type $f(R) = R-\frac{\alpha}{R^n}$, encounter challenges in successfully passing the solar system tests \cite{Chiba:2003ir} and are susceptible to gravitational instabilities \cite{Dolgov:2003px}. Furthermore, these theories fail to generate the standard matter-dominated era preceding accelerated expansion \cite{Amendola:2006kh, Amendola:2006we}. On the other hand, the $f(R) = R + \frac{\alpha}{R^n} + \frac{\beta}{R^m}$ types of models face difficulties in meeting the range of constraints imposed by early and late-time acceleration, big bang nucleosynthesis, and fifth-force experiments \cite{Brookfield:2006mq}.


Considering the numerous challenges associated with most $f(R)$ models mentioned above, more complicated theories were explored, particularly those with an unconventional curvature coupling with the energy-momentum tensor. A family of modified gravity theories that contain the general function $ f(R,T) $ in the gravitational action, where $ R $ and $ T $ stand for the curvature and trace of the energy-momentum tensor, respectively, has recently been presented \cite{Harko:2011kv}. Since the inception of this theory, extensive research has been conducted on various facets, including the thermodynamic properties \cite{Houndjo:2012hj, Sharif:2012zzd, Sharif:2013ffa}, energy conditions \cite{Alvarenga:2012bt, Sharif:2013kga}, cosmological solutions derived from analyzing a homogeneous and isotropic spacetime using phase-space analysis \cite{Shabani:2013djy}, anisotropic cosmology \cite{Kiran:2013dwa, Sharif:2014cpa}, cosmological solutions involving an auxiliary scalar field \cite{Houndjo:2011tu} and investigation of additional aspects \cite{Reddy:2013rea, Baffou:2013dpa, Moraes:2014cxa}. Studies on the cosmological evolution of $f(R,T)$ theory have taken into account future singularities and the reconstruction of cosmological solutions \cite{ Nojiri:2017ncd, Houndjo:2011fb, Jamil:2011ptc}. However, such a tight coupling between the curvature and the trace $ T $ implies a violation of the common continuity equation, a problem that can be resolved by a suitable function $ f(R,T) $ as first demonstrated in Ref \cite{Alvarenga:2013syu}. 

It is commonly understood that modified gravity theories can be classified into two distinct categories depending on the relationship between matter and geometry: minimal coupling and non-minimal coupling. Shabani and colleagues \cite{Shabani:2013djy} conducted a dynamical systems analysis on various Lagrangians in modified gravity, namely a minimal $g(R) + h(T)$ Lagrangian, a pure non-minimal $g(R)h(T)$ Lagrangian, and a non-minimal $g(R)(1+h(T))$ Lagrangian, using the FLRW metric as a background. Additionally, numerous studies explore the cosmological consequences of the non-minimal coupling of matter and geometry in $f(R,T)$ gravity \cite{Haghani:2013oma, Odintsov:2013iba, Sharif:2013kga}. Furthermore, several studies in the literature focus on specific linear versions of $f(R, T)$ gravity \cite{Alvarenga:2012bt, Shabani:2014xvi, Singh:2018xjv, Shabani:2016lnz, Singh:2022jue}. 
Our study focuses explicitly on $f(R,T)$ gravity with minimal coupling, where the function is represented as $f(R,T)=R+2\lambda T$, with $\lambda$ representing an arbitrary coupling constant. 

The following is the presentation and structure of the paper: Section \ref{Sec2} provides a brief overview of the formulation of the field equations and its solution in $f(R, T)$ gravity. In this section, by employing an appropriate parameterization, expressed as a differential equation involving the energy density of the scalar field, $\rho_{\phi}$, we derive equations that characterize the Hubble and deceleration parameters in terms of the redshift. Section \ref{Sec3} provides a concise overview of the observational Hubble, Pantheon, and BAO data utilized to assess the cosmological model and establish constraints on the model parameters. In this context, we aim to demonstrate the close correspondence between observational and theoretical outcomes. In Section \ref{Sec4}, we deliberate the outcomes obtained from the observational data and observe the universe's evolution. This section examines the jerk and snap parameters, revealing that our current model aligns with a dark energy quintessence model.
Additionally, we have depicted the behaviour of the densities and pressure along with the Equation of State (EoS) parameter. Our explanation of energy conditions elucidates the viability of our model. Finally, we study the thermodynamical behaviour of our model in Section \ref{Sec5}, and conclude our work in Section \ref{Sec6}. 

\section{Field Equations}\label{Sec2}
Among the modified theories based on non-minimal interaction between matter and geometry, $ f(R,T) $ gravity is an intriguing contender. This updated theory's behaviour takes the form 
\begin{equation}\label{1}
S=\frac{1}{\kappa^2}\int[f(R,T)+\mathcal{L}]\sqrt{-g}dx^4. 
\end{equation}
Here $ \kappa^2 =1 $ is the gravitational coupling constant, $ R $ is the Ricci scalar, $T$ is the trace of the stress energy-momentum tensor. Also, $ \mathcal{L}=\mathcal{L}_m+\mathcal{L}_{\phi} $ where $ \mathcal{L}_m $ is the matter Lagrangian and $ \mathcal{L}_{\phi} $ is the action for the scalar field. The action of the scalar field is given by
\begin{equation}\label{2}
\mathcal{L}_{\phi}={\Big[-\frac{1}{2}\partial_\mu \phi \partial^\mu \phi -V(\phi)\Big]\sqrt{-g}d^4x}.
\end{equation}
We analyze the evolution and dynamics of the universe for a particular form of $ f(R,T) $, namely $ f=f_1(R)+f_2(T)$ where $f_1(R)=R$ and $f_2(T)=2\lambda T$, $\lambda$ being an arbitrary coupling constant. We assume that the universe is isotropic and homogeneous, with the energy density, $ \rho $, 
the thermodynamic pressure, $ p $, and the matter Lagrangian $\mathcal{L}_m = -\rho_m $. The Friedmann-Lema$\hat{\imath}$tre-Robertson-Walker (FLRW) metric, which \textbf{describes} the geometry of space-time, is given by
\begin{equation}\label{3}
ds^{2}=dt^{2}-a^{2}(t)(dx^2+dy^2+dz^2),   
\end{equation}
where $ a(t) $ is the scale factor of the universe. 
The gravitational field equations in $ f(R,T) $ gravity can be found by varying the action given by (\ref{1}) with respect to $ g_{\mu\nu} $ as
\begin{equation}\label{4}
G_{\mu\nu}+(g_{\mu\nu}\square - \nabla_{\mu}\nabla_{\nu})=[8\pi+2f'(T)]T_{\mu\nu}+2[f'(T)p+\frac{1}{2}f(T)]g_{\mu\nu}.
\end{equation}
The energy-momentum tensor is given by \cite{Landau:1975pou}
\begin{equation}\label{5}
T_{\mu\nu}= -\frac{2}{\sqrt{-g}} \frac{\delta(\sqrt{-g}\mathcal{L})}{\delta g^{\mu\nu}}.
\end{equation}
By assuming a matter action that solely depends on the metric and not on its first derivatives, the energy-momentum tensor thus produces, 
\begin{equation}\label{6}
T_{\mu\nu}= g_{\mu\nu}\mathcal{L}-2\frac{\delta \mathcal{L}}{\delta g^{\mu\nu}}.
\end{equation}
The independent cosmological field equations for the FLRW metric are
\begin{equation}\label{7}
3H^2=(8\pi+3\lambda)\rho_{total}-\lambda p_{total},
\end{equation}
\begin{equation}\label{8}
2\dot{H}+3H^2=-(8\pi+3\lambda)p_{total}+\lambda \rho_{total},
\end{equation}
where we take $ 8\pi=1 $, $\rho_{total}= \rho_m+\rho_{\phi}$, $p_{total}=p_{\phi} $ and a dot indicates differentiation \textit{w.r.t.} time $t$. $\rho_m$, $\rho_{\phi}$, and $p_{\phi}$ denote the energy density of matter, energy density of scalar field, and pressure of scalar field, respectively. 
The isotropy and homogeneity of the model can be explained by the dominant component of the scalar field $\phi$, which only varies with time. Since the scalar field $\phi$ is time-dependent, it can be treated as a perfect fluid with energy density, $\rho_{\phi}$, and pressure, $p_{\phi}$. Assuming that the scalar field, $\phi$, is the sole source of DE, with a potential $V(\phi)$ that interacts with itself, we can consider energy density, $\rho_{\phi}$, and pressure, $p_{\phi}$, as the canonical components of the scalar field $\phi$, within the framework of FLRW cosmology:
\begin{equation}\label{9}
    \rho_{\phi}=\frac{\dot{\phi}^2}{2}+V(\phi) ,\qquad 
    p_{\phi}=\frac{\dot{\phi}^2}{2}-V(\phi).
\end{equation}

In general relativity, the conservation of the energy-momentum tensor is expressed as follows \cite{Singh:2019fpr}:
\begin{equation}\label{10}
\dot{\rho}_m+3\bigg(\frac{\dot{a}}{a}\bigg)\rho_m=0,
\end{equation}
\begin{equation}\label{11}
\dot{\rho}_{\phi}+3\bigg(\frac{\dot{a}}{a}\bigg)(\rho_{\phi}+p_{\phi})=0.
\end{equation}
Eq. (\ref{10}) yields
\begin{equation}\label{12}
\rho_m=\rho_{m_0}a^{-3},
\end{equation}
where $\rho_{m_0}$ is an arbitrary integration constant. Solving Eq. (\ref{11}), we get
\begin{equation}\label{13}
\dot{\rho}_{\phi}=-3\frac{\dot{a}}{a}(1+\omega_{\phi})\rho_{\phi}.
\end{equation}
The EoS parameter of the scalar field $\phi$ is expressed as $\omega_{\phi}=\frac{p_{\phi}}{\rho_{\phi}}$, and its value can be determined using Eq. (\ref{11}) as
\begin{equation}\label{14}
\omega_{\phi}=-1-\frac{1}{3}a\frac{\rho_{\phi}'}{\rho_{\phi}},
\end{equation}
where a prime denotes the derivative \textit{w.r.t.} scale factor $ a $. The system of Eqs. (\ref{7}), (\ref{8}), (\ref{10}) and (\ref{11}) only has three independent equations and four unknowns: $ H$, $\rho_m$, $\rho_{\phi} $ and $ p_{\phi} $. It needs some additional constraint equations to be solved. 

In Eq. (\ref{14}), we find that the EoS parameter $ \omega_{\phi} $ is a function of $ \rho_{\phi}'  $ and the scale factor $ a $; therefore we assume a suitable parametrization in the form of a differential equation in $ \rho_{\phi} $ as 
\begin{equation}\label{15}
\rho_{\phi}'+\alpha f(a) \rho_{\phi}=0,
\end{equation}
where $ f(a)=\Big[1-\frac{1}{\alpha(1+a)\log(1+a)}\Big] $.

The function $ f(a) $ containing the term $ \Big[1-\frac{1}{\alpha(1+a)\log(1+a)}\Big] $ has been considered in such a way that our model can show the consistency with present observational datasets within the framework of the modified theory of gravity and exhibits accelerated expansion in later times. Various parametrizations in different mathematical forms of cosmological parameters have already been used in cosmology to study dark energy models \cite{Singh:2019fpr, Cunha:2008ja, Linder:2002et}. These parametrizations neither presume the validity of any gravitational theory nor affect the model by violating any physical and geometrical properties \cite{Wein}.

The generic answer to the differential Eq. (\ref{15}) mentioned above is given by
\begin{equation}\label{16}
\rho_{\phi}(z)=e^{-\alpha a}\log(1+a),
\end{equation}
where $\alpha \in (0,1)$ is the model parameter. 
The energy density of scalar field $\rho_{\phi}$ can be expressed in terms of redshift $z$ by using the relationship between redshift $z$ and scale factor $a$, which is given by $\frac{a}{a_0}=\frac{1}{1+z}$, where $a_0 = 1$ represents the current value of the scale factor
\begin{equation}\label{17}
\rho_{\phi}(z)=e^{\frac{-\alpha }{1+z}}\log\bigg(1+\frac{1}{1+z}\bigg),
\end{equation}
and
\begin{equation}\label{18}
\rho_{\phi_0}=e^{-\alpha}\log(2).
\end{equation}
Eqs. (\ref{17}) and (\ref{18}) yield
\begin{equation}\label{19}
  \rho_{\phi}(z)= \frac{\rho_{\phi_0}}{\log(2)}e^{\frac{\alpha z}{1+z}}\log\bigg(1+\frac{1}{1+z}\bigg).  
\end{equation}
where $\rho_{\phi_0}$ is the current value of the energy density of the scalar field. Using Eq. (\ref{12}), it is possible to express the energy density of the matter field $ \rho_m $ in terms of redshift $ z $ as
\begin{equation}\label{20}
\rho_m(z)=\rho_{m_0}(1+z)^3.
\end{equation}
Combining Eqs. (\ref{7}), (\ref{17}) and (\ref{19}) we obtain 
\begin{multline}\label{21}
3 H(z)^2=(3 \lambda +1) \left((z+1)^3 \rho _{\text{m0}}+\frac{\rho _{\text{$\phi $0}} e^{\frac{\alpha  z}{z+1}} \log \left(\frac{1}{z+1}+1\right)}{\log (2)}\right)\\-\lambda  \left(-\frac{\frac{\rho _{\text{$\phi $0}} e^{\frac{\alpha  z}{z+1}} \left(\frac{\alpha }{z+1}-\frac{\alpha  z}{(z+1)^2}\right) \log \left(\frac{1}{z+1}+1\right)}{\log (2)}- \frac{\rho _{\text{$\phi $0}} e^{\frac{\alpha  z}{z+1}}}{(z+1)^2 \left(\frac{1}{z+1}+1\right) \log (2)}}{3 (z+1)}-\frac{\rho _{\text{$\phi $0}} e^{\frac{\alpha  z}{z+1}} \log \left(\frac{1}{z+1}+1\right)}{\log (2)}\right).
\end{multline}
The density parameter $\Omega = \rho/\rho_c$ describes the universe's contents, with $\rho_c =3H^2/(8\pi G)^2$ representing the critical density. In normalized units $8 \pi G = 1$. We can express Eq. (\ref{21}) in terms of the density parameter of matter and the scalar field as
\begin{multline}\label{22}
 H(z)^2=H_0^2(3 \lambda +1) \left((z+1)^3 \Omega _{\text{m0}}+\frac{\Omega _{\text{$\phi $0}} e^{\frac{\alpha  z}{z+1}} \log \left(\frac{1}{z+1}+1\right)}{\log (2)}\right)\\- H_0^2\lambda  \left(-\frac{\frac{\Omega _{\text{$\phi $0}} e^{\frac{\alpha  z}{z+1}} \left(\frac{\alpha }{z+1}-\frac{\alpha  z}{(z+1)^2}\right) \log \left(\frac{1}{z+1}+1\right)}{\log (2)}- \frac{\Omega _{\text{$\phi $0}} e^{\frac{\alpha  z}{z+1}}}{(z+1)^2 \left(\frac{1}{z+1}+1\right) \log (2)}}{3 (z+1)}-\frac{\Omega _{\text{$\phi $0}} e^{\frac{\alpha  z}{z+1}} \log \left(\frac{1}{z+1}+1\right)}{\log (2)}\right),
\end{multline}
where $\Omega_{m_0}=\frac{\rho_{m_0}}{3H_0^2}$ and $ \Omega_{\phi_0}=\frac{\rho_{\phi_0}}{3H_0^2}$ are the present values of matter and scalar field density parameters respectively. We can write

\begin{multline}\label{23}
    H= \frac{1}{(z+1)^{3/2} \sqrt{z+2} \sqrt{\log (8)}} \Bigg(\surd \Big(H_0^2(4.28365 + z(27.8437 + z (77.1057 + z (117.8 + z(107.091 + z (57.8293 + z \\(17.1346 + 2.14182 z)))))))\Big)\Omega_{m_0}+\Omega_{\phi_0}(-0.01 z-0.01) e^{\frac{\alpha  z}{z+1}}\\
    + e^{\frac{\alpha z}{1+z}}(0.06 + H_0^2(6.18 + z (21.63 + z(27.81 + z (15.45 + 3.09 z)))))\\
    +0.02 \alpha +z (0.01 \alpha +z ((0.03 z+0.15) z+0.27)+0.21))\Omega_{\phi_0}\log \left(\frac{1}{z+1}+1\right)\Bigg).
\end{multline}
The following section involves constraining the model parameters $\Omega_{m_0},\Omega_{\phi_0}, H_0$, and $\alpha$ by utilizing observational datasets and determining their optimal fit values. Subsequently, these values will be utilized to analyze the behaviour of various physical parameters.

\section{Observation Dataset}\label{Sec3}
In the previous section, we examined $f(R,T)$ gravity and found an exact solution for the resulting field equations. The solution includes four model parameters, namely,  $\Omega_{m0},\Omega_{\phi 0},H_0,$ and $\alpha$. To verify our method, we need to limit these model parameters using observational datasets to determine their best-fit values. Three datasets were used in this study: the Hubble dataset, which includes 77 data points, the recently released Pantheon sample, which includes 1048 Supernova Type Ia experiment findings from surveys such as the Low-z, SDSS, SNLS, Pan- STARRS1(PS1) Medium Deep Survey, and HST \cite{Pan-STARRS1:2017jku} in the redshift range $z \in (0.01, 2.26)$ and the BAO dataset consisting of 6 data points. 
\begin{figure}[ht]
\includegraphics[scale=0.45 ]{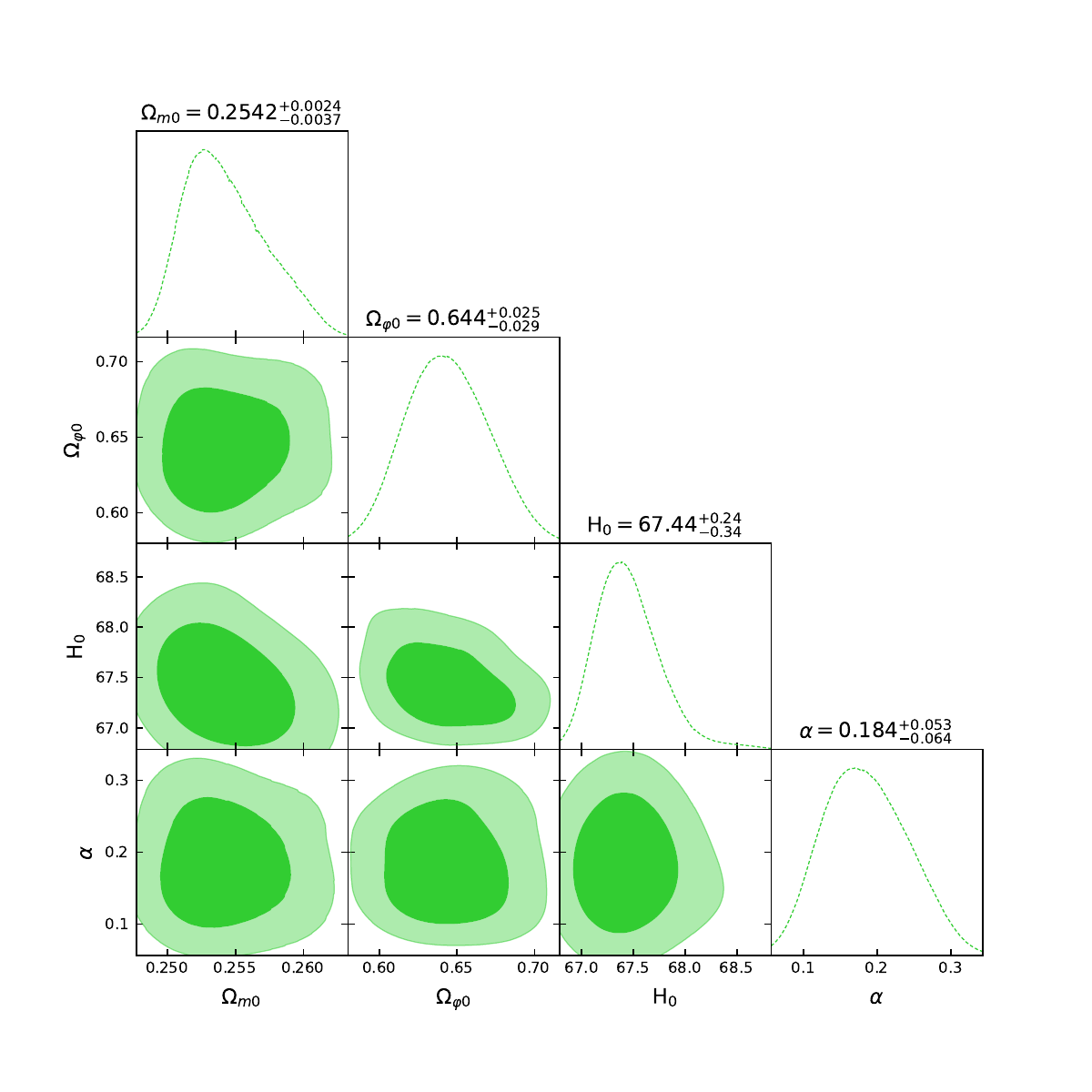}
	\caption{ The graph provides the best-fit values of the model parameters $\Omega_{m0}$, $\Omega_{\phi 0}$, $H_0,$ and $ \alpha $ with $1-\sigma$ and $2-\sigma$ errors for the 77-point Hubble dataset.}
 \label{Hub}
\end{figure}

\subsection{Hubble Data}
Observational cosmology employs the Hubble parameter to examine the expansion of the universe. Two widely used techniques for estimating $H(z)$ at certain redshifts are the extraction of $H(z)$ from Baryon Acoustic Oscillation (BAO) data and the differential age method. Ref. \cite{Shaily:2022enj}  provides a summary of the revised dataset containing 77 points in the redshift range of 
$ 0.07 < z < 2.42 $. We calculate the model parameters, we utilize MCMC simulation and the $\chi^2$ test. The value of $\chi^2_{Hubble} $ for the observational Hubble parameter can be computed as  \begin{equation}\label{24}
    \chi^2_{Hubble}=\sum_i^{77}\frac{[H_{th}(z_i,\Omega_{m0},\Omega_{\phi 0},H_0,\alpha)-H_{obs}(z_i)]^2}{\sigma^2_{(z_i)}}.
\end{equation}
Here, the theoretical value of the Hubble parameter is denoted by $H_{th}$, while $H_{obs}$ denotes the observed value. The standard error of the observed Hubble parameter is represented by $\sigma_{(z_i)}$.

Using the Hubble dataset consisting of $77$ points, as mentioned earlier, we generated two-dimensional contour plots with $1-\sigma$ and $2-\sigma$ errors in Fig. \ref{Hub} to determine the optimal values of the model parameters $\Omega_{m0}$, $\Omega_{\phi 0}$, $H_0,$ and $ \alpha $. The best fit values are obtained as $ \Omega_{m0}=0.2542^{+0.0024}_{-0.0037}$, $\Omega_{\phi 0}= 0.644^{+0.025}_{-0.029}$, $H_0=67.44^{+0.24}_{-0.34}$ and $\alpha=0.184^{+0.053}_{-0.065}$.

\subsection{Pantheon Data}
\begin{figure}[ht]
\includegraphics[scale=0.42]{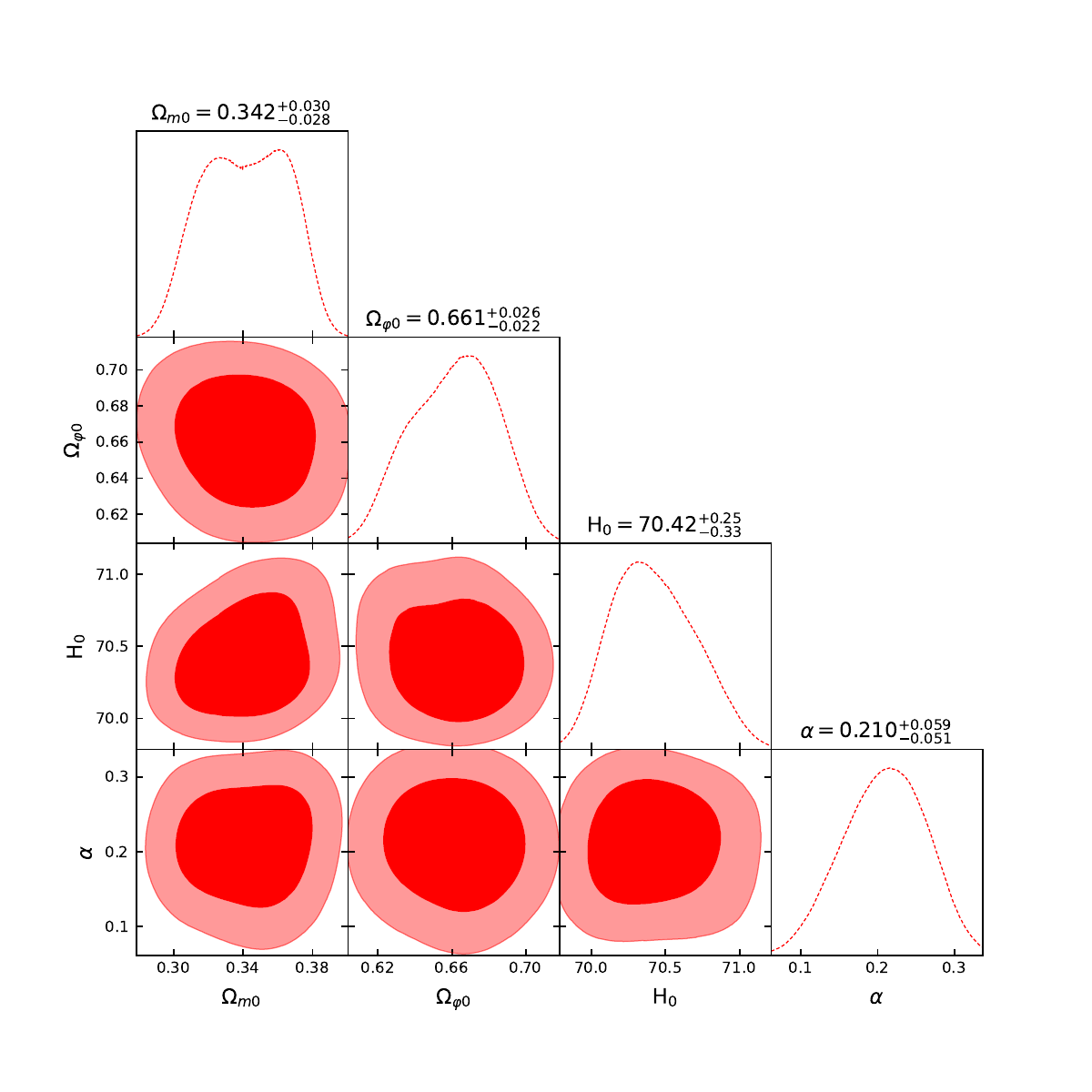}
	\caption{ The plot depicts the best-fit values of the model parameters $\Omega_{m0}$, $\Omega_{\phi 0}$, $H_0,$ and $ \alpha $ with $1-\sigma$ and $2-\sigma$ errors for the $1048$ pantheon dataset.}
 \label{Pan}
\end{figure}

The most recently published supernovae type $Ia$ dataset is the Pantheon sample, consisting of 1048 data points. In our analysis, we utilize these data points, confirmed spectroscopically by SNeIa, and cover the redshift range of $0.01 < z < 2.26$.  The $\chi^2_{Pantheon}$ function for the Pantheon dataset is taken to be,
\begin{equation}\label{25}
    \chi_{Pantheon}^{2}=\sum\limits_{i=1}^{1048}\left[ \frac{\mu_{th}(\mu_0,\Omega_{m0},\Omega_{\phi 0},H_0,\alpha)-\mu_{obs}(z_{i})}{\sigma _{\mu(z_{i})}}\right] ^2.
\end{equation}
In addition, the symbol $\sigma_{\mu(z_i)}$ represents the standard error in the actual value of $\mu$. The distance modulus $\mu_{th}$ is defined as $\mu^i_{th} = \mu(D_L)= m - M = 5log_{10}D_L(z) +\mu_0$, where $m$ and $M$ represent the apparent and absolute magnitudes, respectively, and the nuisance parameter $\mu_0$ is determined by $\mu_0 = 5log(H_0^{-1}/Mpc ) + 25$.
To obtain the luminosity distance $D_L$, one can utilize the formula $D_L(z)=(1+z)H_0\int \frac{1}{H(z^*)}dz^*$, where $H(z)$ is a series that is limited to the tenth term and can be roughly integrated to determine the luminosity distance. The most suitable values of the model parameters $ \Omega_{m0}$, $\Omega_{\phi 0}$, $H_0$ and $\alpha$ have been established by examining two-dimensional contour plots with $1-\sigma$ and $2-\sigma$ errors, using the Pantheon dataset mentioned above. The resulting best-fit values based on the $1048$ Pantheon dataset are $ \Omega_{m0}=0.342^{+0.030}_{-0.028}$ , $\Omega_{\phi 0}= 0.661^{+0.026}_{-0.022}$, $H_0=70.42^{+0.25}_{-0.33}$ and $\alpha=0.210^{+0.059}_{-0.051}$ as shown in Fig. \ref{Pan}.

\subsection{BAO Dataset}
\begin{figure}[ht]\centering
	\subfloat[]{\label{a}\includegraphics[scale=0.35]{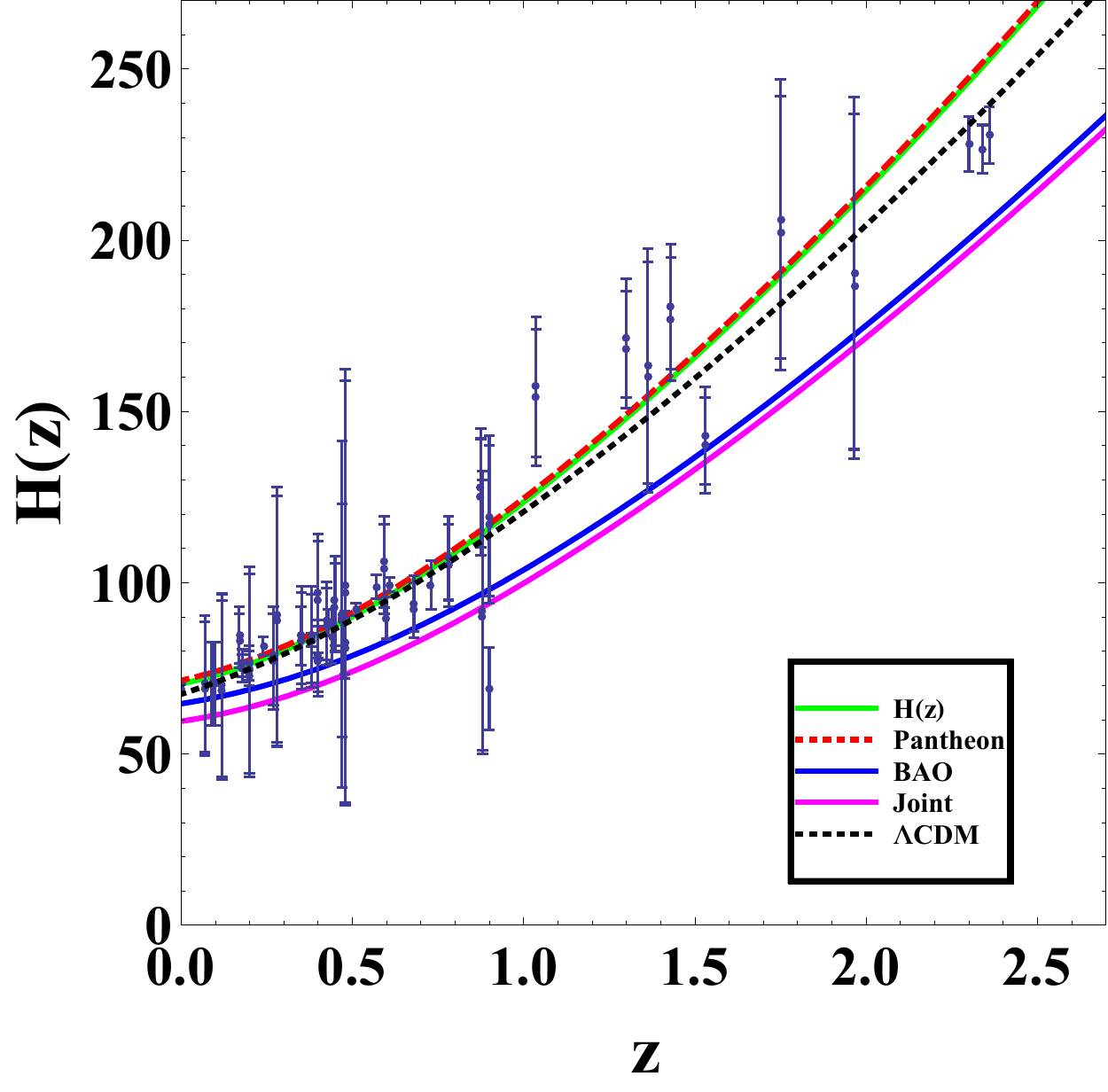}}\hfill
	\subfloat[]{\label{b}\includegraphics[scale=0.34]{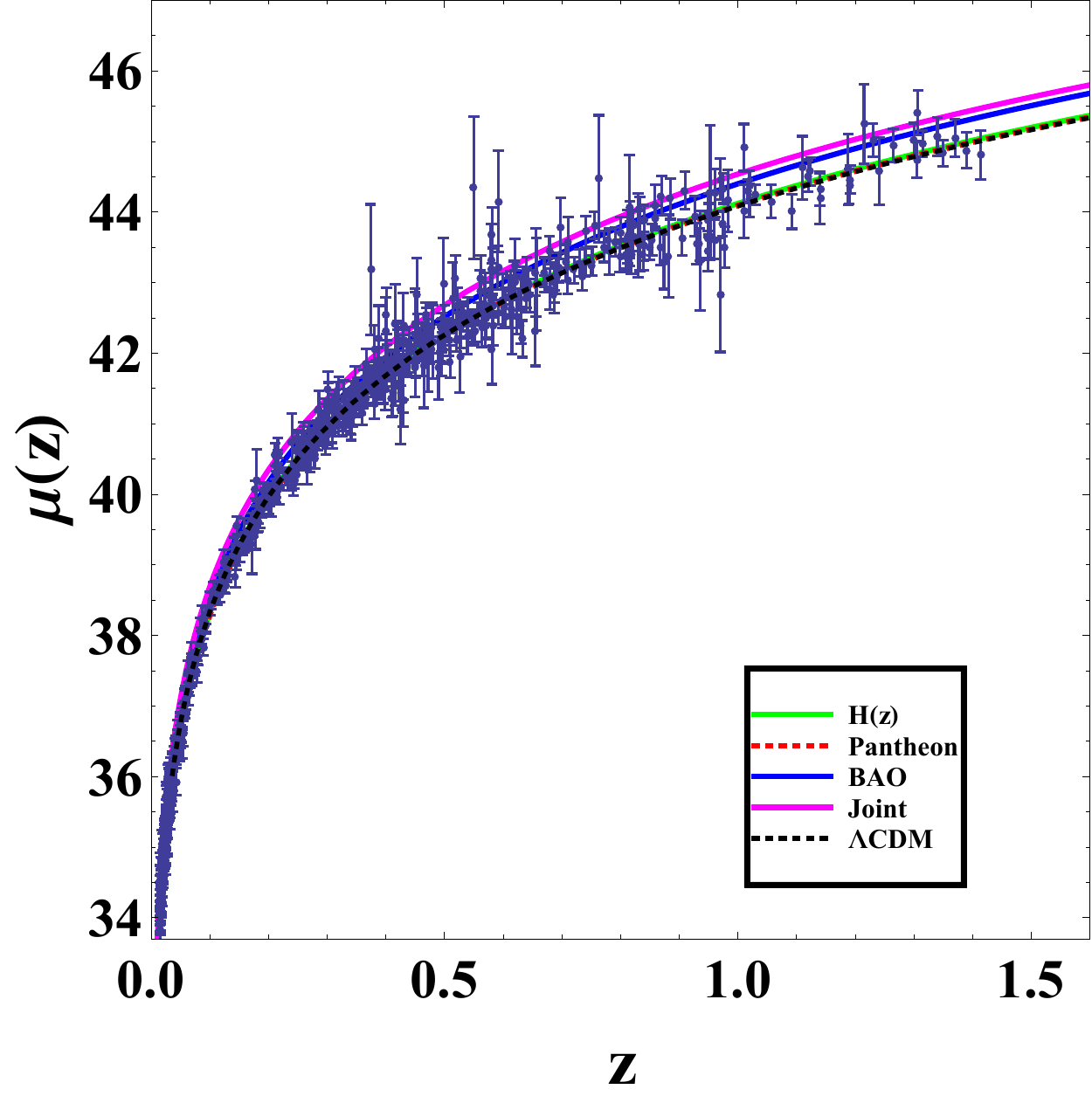}} 
	\caption{ The error bar plots show the variations of the Hubble parameter $ H(z) $ and the distance modulus $\mu(z)$ in comparison to the $\Lambda$CDM.}
 \label{H-mu}
\end{figure}

BAO represents a significant cosmological event that emerged during the universe's infancy. This BAO scale acts as a universal benchmark within the vast structure of the cosmos, offering an unparalleled and resilient cosmic measuring tool for gauging the historical expansion of the universe. In this study, we have incorporated six points of the Baryon Acoustic Oscillation (BAO) dataset, sourced from various surveys. These datasets encompass observations from the Six Degree Field Galaxy Survey (6dFGS), the Sloan Digital Sky Survey (SDSS), and the LOWZ samples of the Baryon Oscillation Spectroscopic Survey (BOSS) \cite{Blake:2011en, SDSS:2009ocz}.
The expression for the scale of dilation, denoted as $D_v(z)$, as presented in \cite{SDSS:2005xqv}, is as follows
\begin{equation}
    D_v(z)=\Big(\frac{d^2_A(z)z}{H(z)}\Big)^{1/3}.
\end{equation}
Here $d_A(z)$ represents the comoving angular diameter distance and is formally defined as follows:
\begin{equation}
    d_A(z)=\int_0^z\frac{dz'}{H(z')},
\end{equation}
The chi-squared function for the BAO analysis is defined as
\begin{equation}
    \chi^2_{BAO}=A^T C^{-1}_{BAO}A.
\end{equation}
A's dependency relies on the specific survey being considered, while $C^{-1}$ represents the inverse of the covariance matrix \cite{{Giostri}}. 
\begin{figure}[ht]
\includegraphics[scale=0.42]{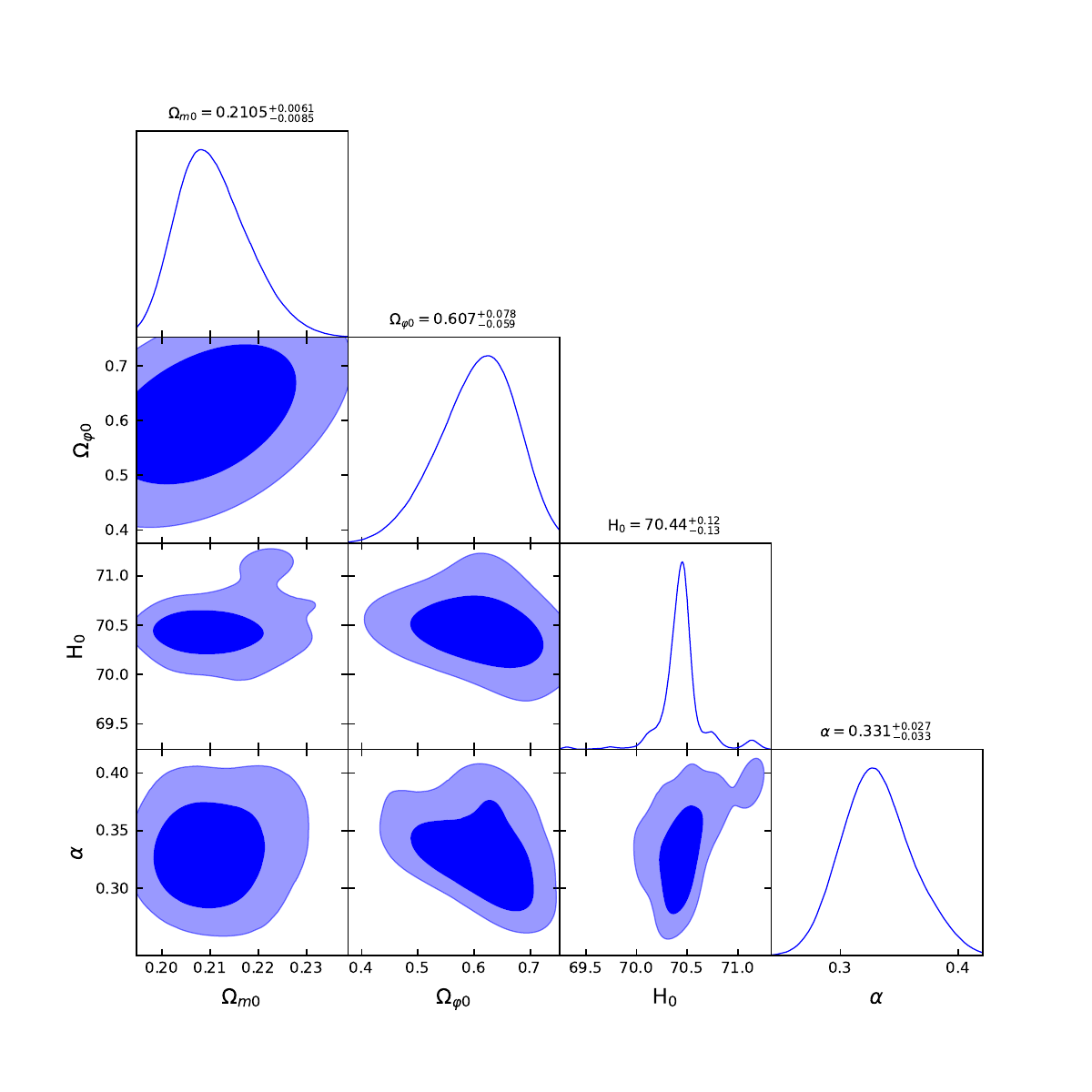}
	\caption{ The graph gives the best-fit values of the model parameters $\Omega_{m0}$, $\Omega_{\phi 0}$, $H_0,$ and $ \alpha $ with $1-\sigma$ and $2-\sigma$ errors for the 6 points of the BAO dataset.}
 \label{BAO}
\end{figure}
The optimal parameter values for the model, including $\Omega_{m0}$, $\Omega_{\phi 0}$, $H_0$, and $\alpha$, have been determined by analyzing two-dimensional contour plots, which incorporate $1-\sigma$ and $2-\sigma$ uncertainties, using the aforementioned BAO dataset. The resulting parameter estimates based on the 6 data points from the BAO dataset are as follows: $\Omega_{m0}=0.2105^{+0.0061}_{-0.0085}$, $\Omega{\phi 0}=0.607^{+0.078}_{-0.059}$, $H_0=70.44^{+0.12}_{-0.13}$, and $\alpha=0.331^{+0.027}_{-0.033}$, as illustrated in Fig. \ref{BAO}.
\begin{figure}[ht]
\includegraphics[scale=0.45]{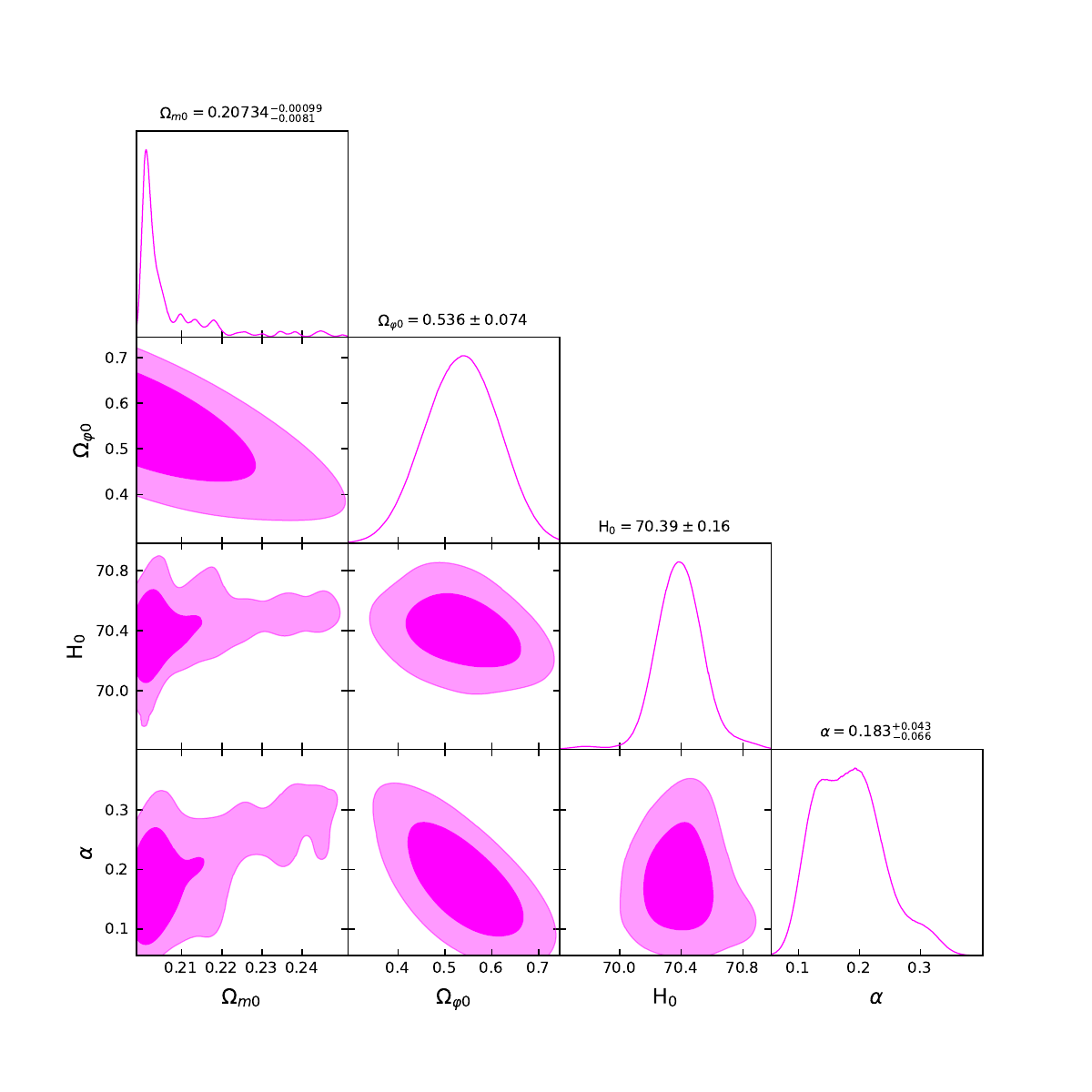}
\caption{ The contour plots for model parameters $\Omega_{m0}$, $\Omega_{\phi 0}$, $H_0,$ and $ \alpha $  with $1 - \sigma$ and $2 - \sigma$ errors which provide the best-fit values of the model parameters from the joint datasets of Hubble, Pantheon, and BAO datasets.}
 \label{Hz+Pan}
\end{figure}
The left panel of Fig. \ref{H-mu} depicts the Hubble function $H(z)$ development over varying redshift $z$, where colourful lines represent the obtained model. The blue bars on the curve display the 77 data points of the Hubble dataset, along with their respective error bars. The black dotted line also illustrates the $ \Lambda $CDM model. The coloured lines in the right panel of Fig. \ref{H-mu} demonstrate the relationship between $\mu(z)$ and $z$ for our model, while the black dotted line represents the curve for the $\Lambda$CDM model. Both models match the $H(z)$ data, Union 2.1 compilation data, and their respective error bars.
Having obtained theoretical formulas and numerical values for the model parameters, we are now equipped to delve into the physical dynamics of the model. Thus, the following section will be focused on exploring the physical dynamics of additional significant cosmological parameters. The constrained values of the model parameters using various observational datasets have been summarised in Table \ref{tabparm1}.

\begin{table}[ht]
\caption{ Summary of the present constrained values of the model parameters using various observational datasets.}\label{tabparm1}
\begin{center}
\begin{tabular}{l c c c c r} 
\hline\hline
{Dataset} &  $ H $ {\footnotesize(km/s/Mpc)} \,\,\,   &  \,\,\,   $ \Om_{m} $ \,\,\, & $ \Om_{\phi} $ \,\,\, & $ \alpha $  \,\,\,
\\
\hline 
\\
{\scriptsize {$ H(z) $ }}  &    $ 67.44^{+0.24}_{-0.34} $    &   $ 0.254^{+0.0024}_{-0.0037} $   &   $ 0.644^{+0.025}_{-0.029} $ & $ 0.184^{+0.053}_{-0.065}$
\\
\\
{\scriptsize { $ \!\! Pantheon $  }}  &  $70.42^{+0.25}_{-0.33} $  &  $ 0.342^{+0.030}_{-0.028} $  &  $ 0.661^{+0.026}_{-0.022} $ & $ 0.210^{+0.059}_{-0.051} $
\\
\\
{\scriptsize { $ BAO $ }} &  $ 70.44^{+0.12}_{-0.13} $  &  $ 0.211^{+0.0061}_{-0.0085} $   &  $ 0.607^{+0.078}_{-0.059} $ & $ 0.331^{+0.027}_{-0.033} $
\\
\\
{\scriptsize {$ H(z) $ + $ Pantheon$ + $ BAO $ } } &  $ 70.39^{+0.16}_{-0.16} $  &  $ 0.207^{-0.0010}_{-0.0081} $  &  $ 0.536^{+0.074}_{-0.074} $ & $0.183^{+0.043}_{-0.066}$
\\
\\
\hline\hline  
\end{tabular}    
\end{center}
\end{table}

\section{ Physical evaluation of the model}\label{Sec4}

\begin{figure}[ht]\centering
	\subfloat[]{{\label{a}\includegraphics[scale=0.35]{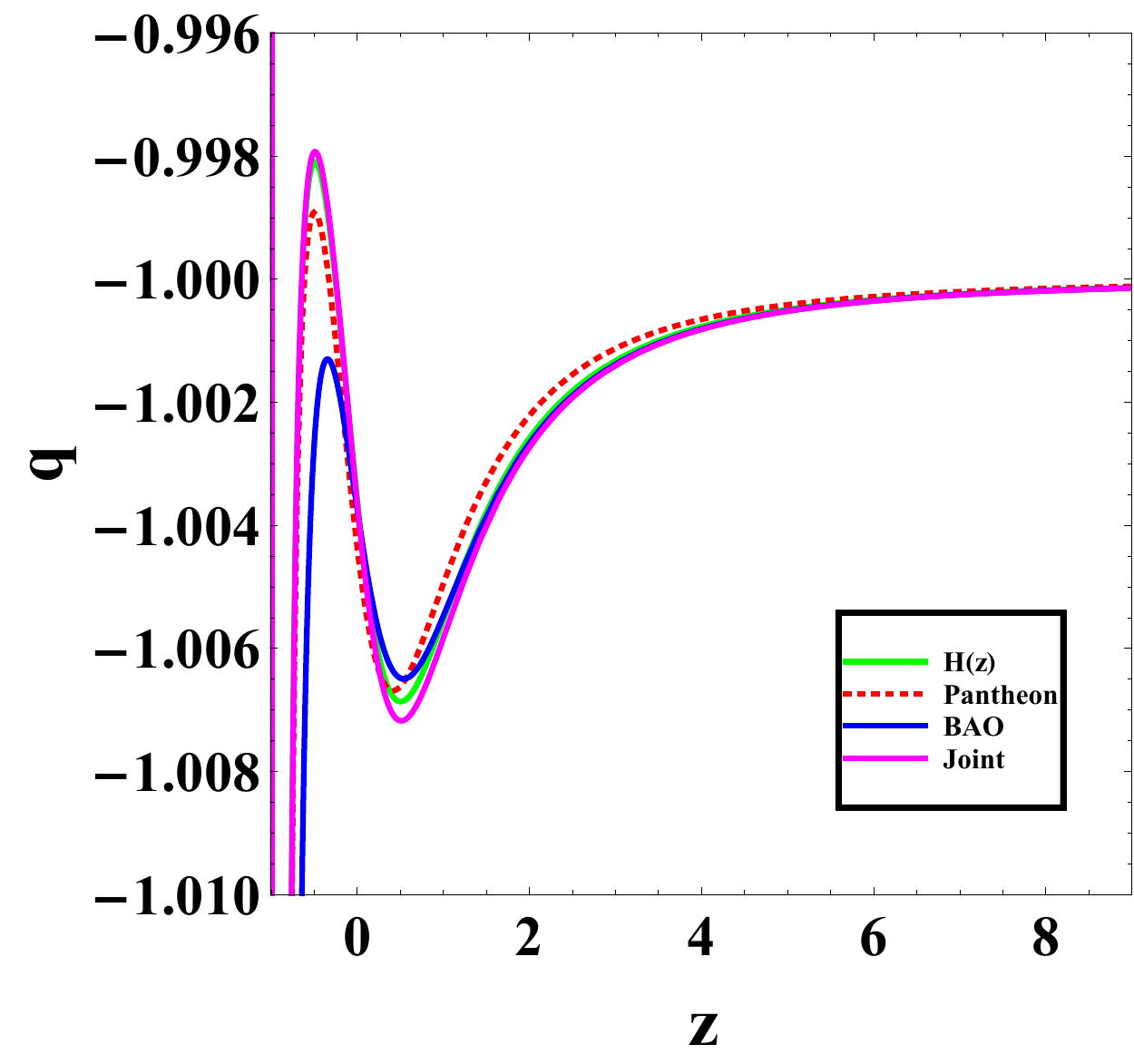}}
 \label{DP}}\hfill
	\subfloat[]{{\label{b}\includegraphics[scale=0.66]{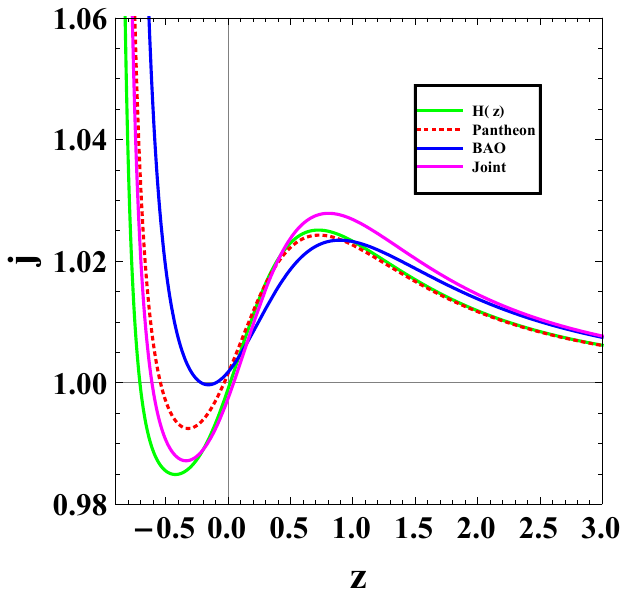}} 
 \label{jerk}}
	\caption{ The behaviour of the deceleration parameter $q$ and jerk parameter $j$ with respect to the redshift $z$. }
 \label{DP-J}
\end{figure}

Numerous dark energy theories have been put forth in the literature to discuss the universe's accelerated expansion. Among various factors that trace the dynamics of the universe, the deceleration parameter (DP), $ q $, is said to be one of the most important cosmic parameters \cite{Bolotin:2015dja}. The rate of acceleration of the universe is described by $q$, which is a function of the Hubble parameter and has the form $q=-1-\frac{\dot{H}}{H^2}$. The universe decelerates if $q$ is positive, while a negative $q$ indicates an accelerated phase. The values of $\Omega_{m0}$, $\Omega_{\phi 0}$, $H_0,$ and $ \alpha $ are utilized in the computation of $q$. Fig. \ref{DP} illustrates the behaviour of $q$ versus redshift $z$ and depicts the expansion from past to present. As is evident, $q$ is in the accelerated stage at present, which can be confirmed from various cosmic observations that show the universe experiences a cosmic acceleration in late periods with a slower pace of expansion \cite{SST:1998fmf, SCP:1998vns}. Negative inflation in the universe is evidenced by the sudden halt in the expansion after the phase transition during late times, approximately at $z = -1$ (refer to Fig. \ref{DP}).

\begin{figure}[ht]\centering
	\subfloat[]{{\label{a}\includegraphics[scale=0.45]{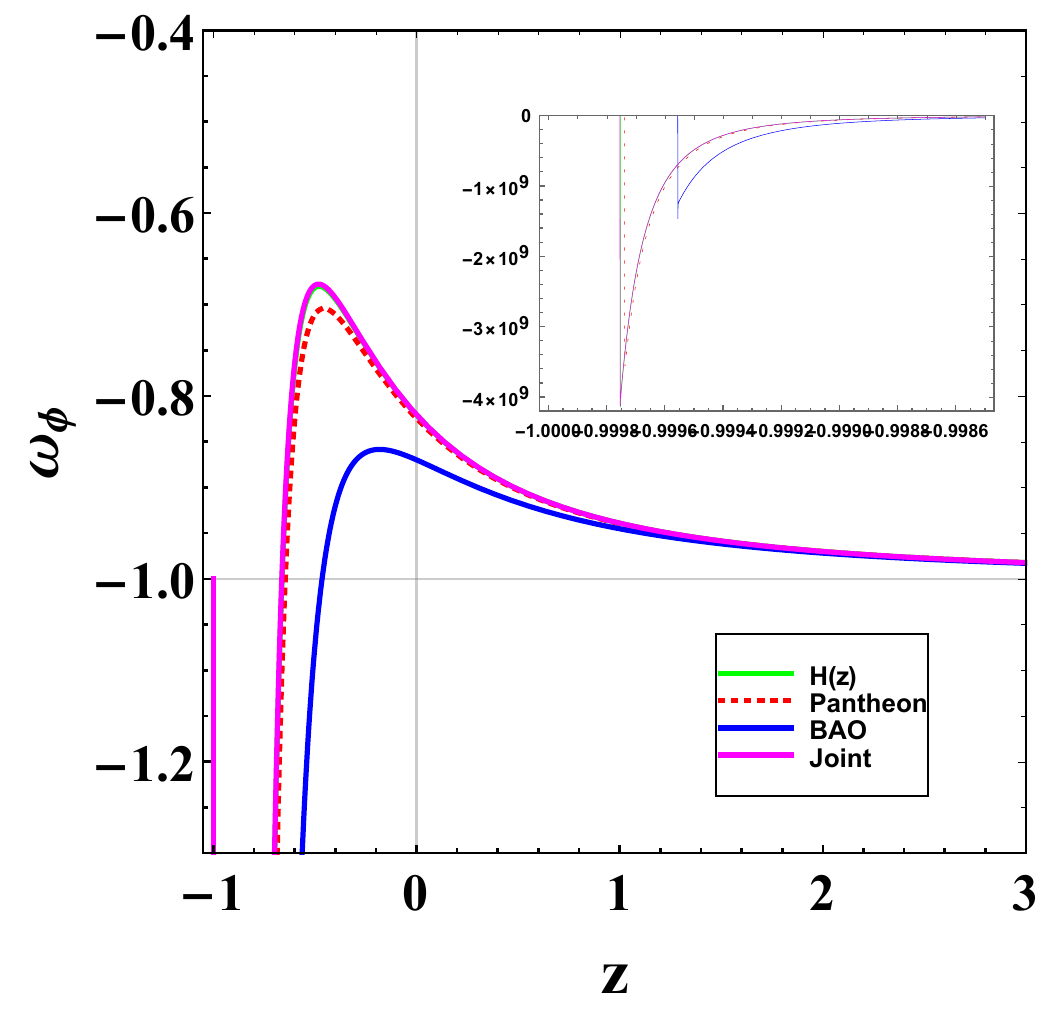}}
\label{wphi}}\hfill
	\subfloat[]{{\label{b}\includegraphics[scale=0.45]{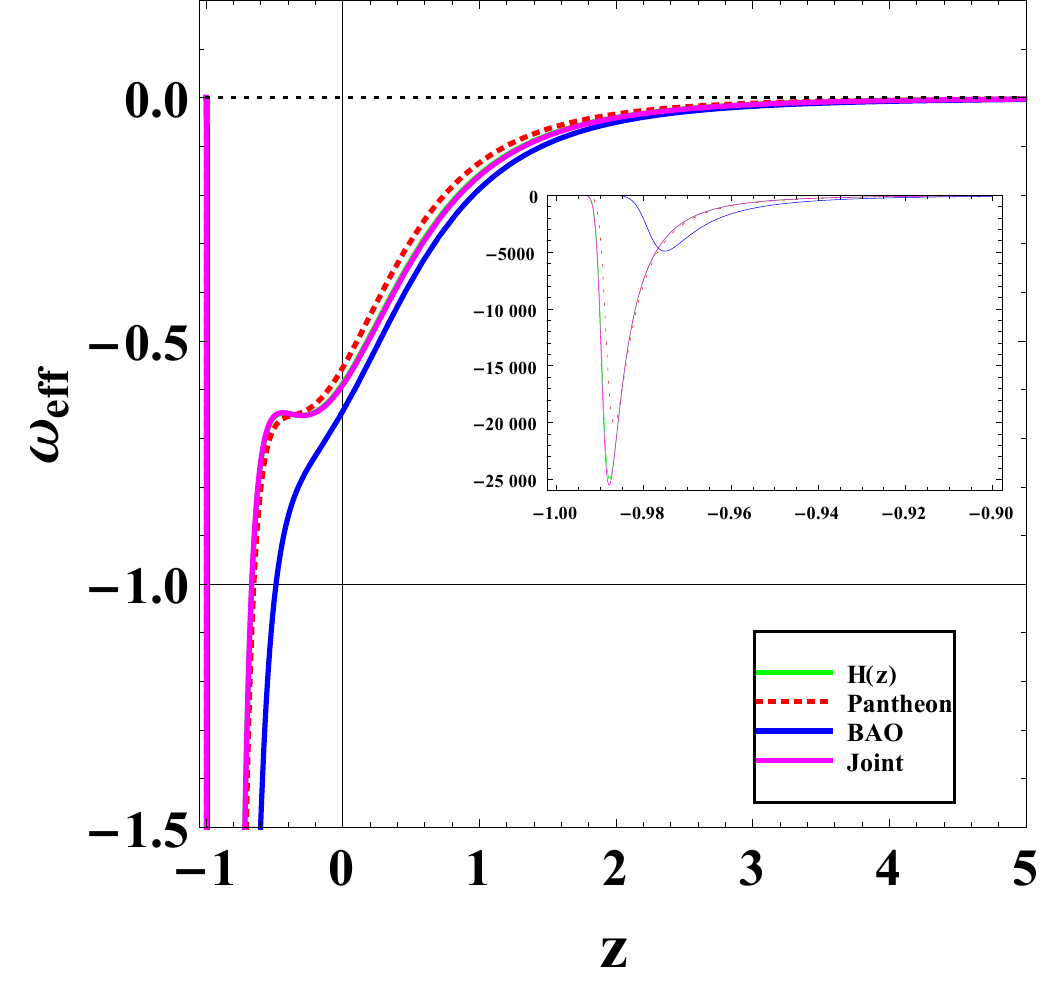}} 
 \label{wtotal}}
	\caption{ Evolution of the EoS parameter $ \omega_{\phi} $ and $\omega_{eff}$ as a function of redshift $z$ for distinct values of model parameters  respectively.}
 \label{omega}
\end{figure}

\begin{figure}[ht]
 \subfloat[]{
	\begin{minipage}[c][1\width]{
	   0.3\textwidth}
	   \centering
	   \includegraphics[width=1.05\textwidth]{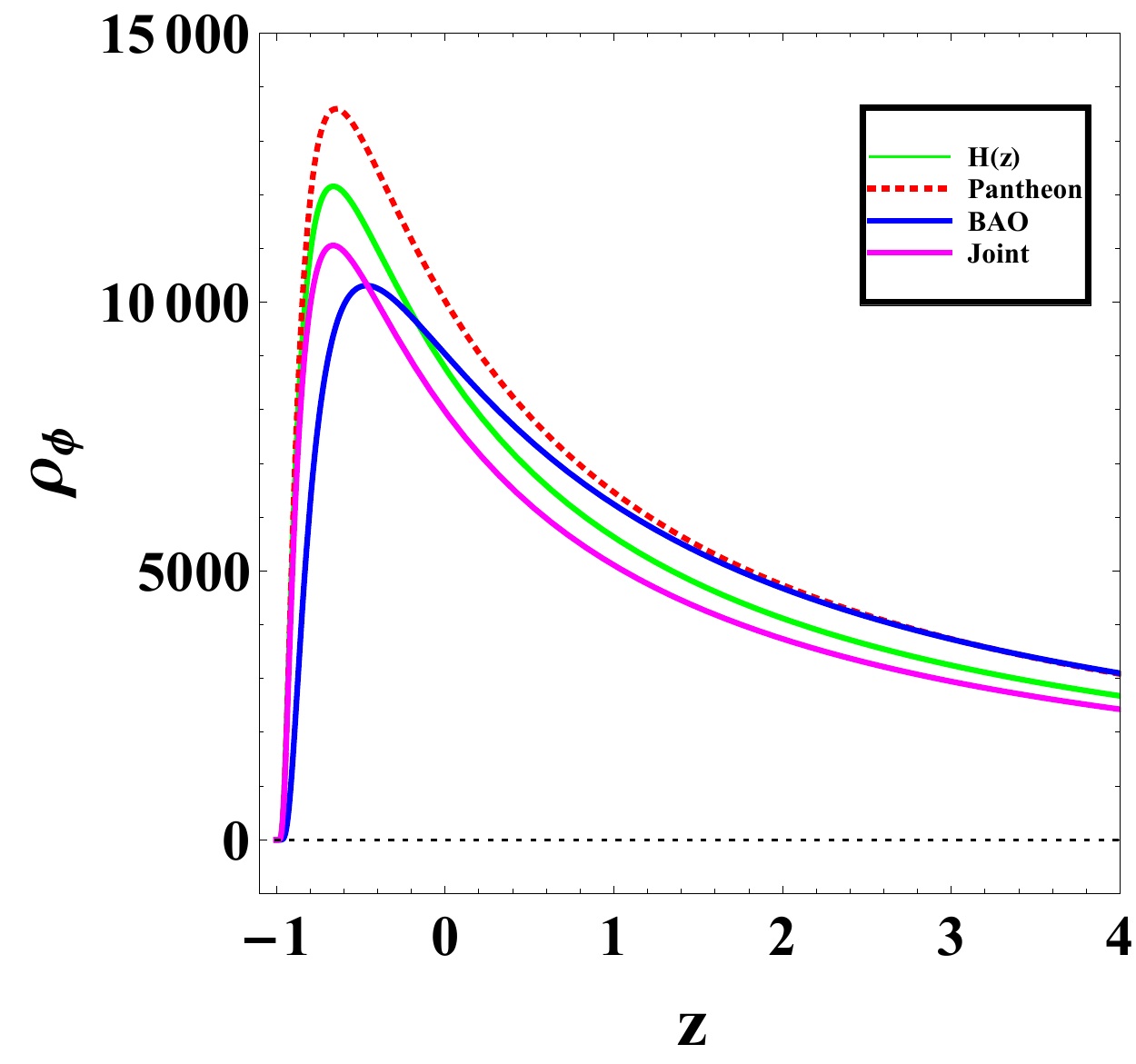}
    \label{rhophi}
	\end{minipage}}
 \hfill 	
  \subfloat[]{
	\begin{minipage}[c][1\width]{
	   0.3\textwidth}
	   \centering
	   \includegraphics[width=1.05\textwidth]{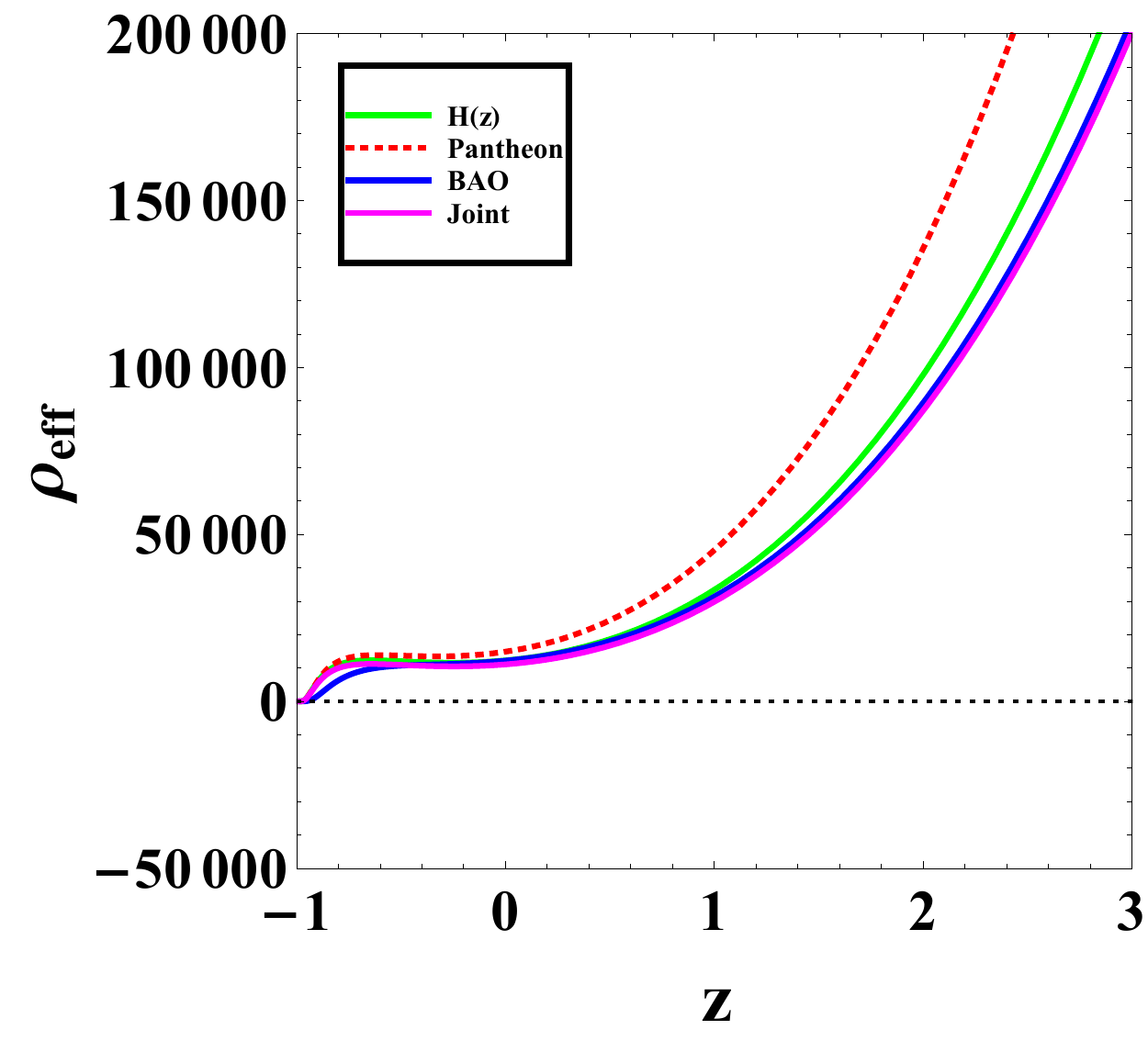}
    \label{rho}
	\end{minipage}}
 \hfill	
 \subfloat[] {
	\begin{minipage}[c][1\width]{
	   0.3\textwidth}
	   \centering
	   \includegraphics[width=1.05\textwidth]{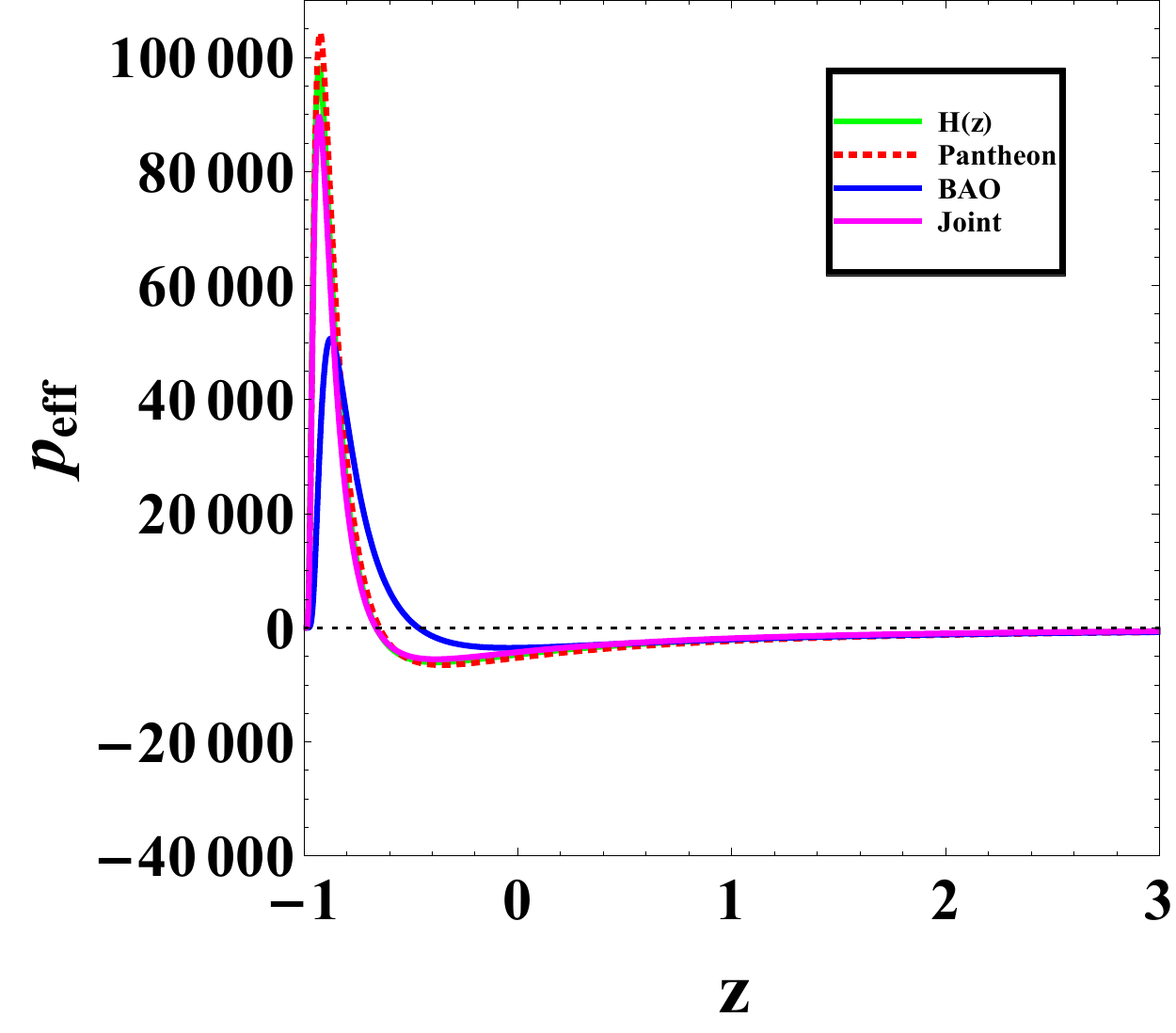}
    \label{pressure}
	\end{minipage}}
\caption{Evolution of $ \rho_{\phi} $, $\rho_{eff}$, and $p_{eff}$ vs redshift $z$ for distinct values of model parameters  respectively.}
\label{rho-p}
\end{figure}

The Hubble, deceleration, and jerk parameters are purely kinematical, which means they are not dependent on any gravity theory. These parameters are solely related to the scale factor $a$ or redshift $z$ (as $a=1/1+z$). The jerk parameter, which is a dimensionless third derivative of the scale factor $a(t)$ \textit{w.r.t.} cosmic time $t$, can be used as a straightforward way to look for deviations from the standard $\Lambda$CDM model. It is defined as \cite{Blandford:2004ah,Rapetti:2006fv,Chiba:1998tc,Visser:2003vq,Luongo:2013rba,AlMamon:2018uby}
\begin{equation}
j=\frac{\dddot{a}}{aH^3}.
\end{equation}
The jerk parameter can be expressed in terms of the deceleration parameter as
\begin{equation}
    j=q(2q+1)+(1+z)\frac{dq}{dz}.
\end{equation}
In their study, Blandford et al. \cite{Blandford:2004ah} explained how the jerk parameter can serve as a convenient and alternate way of describing cosmological models similar to the concordance $ \Lambda $CDM model. The significant advantage of using $j$ is that, for the $\Lambda$CDM model, $j$ always remains constant and equal to $1$. It is worth mentioning that Sahni et al. \cite{Sahni:2002fz, Alam:2003sc} emphasized the significance of $j$ in distinguishing between various dark energy models, as any deviation from $ j=1 $ would indicate a preference for a non-$\Lambda$CDM model. Fig. \ref{jerk} displays the model's evolution of the jerk parameter, revealing that for all numerically restricted values of the model parameters $\Omega_{m0},\Omega_{\phi 0}, H_0,$ and $\alpha$, the value of $j_0$ for our model falls within the range of 0.99 to 1.001. A value of approximately $1$ for the cosmic jerk parameter at redshift $z = 0$ indicates the occurrence of cosmic acceleration.
In the study of cosmological models, the kinematic variables hold considerable importance.  
Likewise, the equation of state (EoS) parameter, $\omega$, helps understand the physical significance of energy sources during the universe's evolution. The EoS parameter is defined as
\begin{equation}\label{26}
    \omega= \frac{p}{\rho}.
\end{equation}
Based on observations, $\Lambda$CDM is a compelling contender for dark energy with $\omega = -1$. Nonetheless, the universe's acceleration stage is only understandable when $\omega < -1/3$. This encompasses the quintessence phase, where $0 < \omega < -1$, while a value below $ -1 $ is referred to as the phantom regime. The EoS parameter of the scalar field $\omega_{\phi}$ and the total EoS parameter $\omega_{eff}$ have been examined in our model. The profile of the EoS parameter of scalar field $\omega_{\phi}$ with redshift $z$ is shown in the left panel of Fig. \ref{omega}. It is noticeable that $\omega_{\phi}$  begins in the quintessence region during the universe's evolution, remains in that region up to the present time, and then experiences a sudden collapse before converging to $\Lambda$CDM in the future. The quintessence phase can be observed in the profile of $\omega_{eff}$ shown in the right panel of Fig. \ref{omega}. It is evident that $\omega_{eff}$ is less than zero, and the present value of $\omega_{eff}$ lies between $-0.53$ and $-0.56$. This finding is consistent with specific investigations \cite{Gong:2006gs}, which suggest a period of acceleration at present. In the future, it experiences a sudden collapse before emerging again.
We plot graphs illustrating the energy density as a function of the redshift $z$, considering various constraint values of the model parameters derived from different observational datasets. As the universe evolves, the energy density is observed to be high initially but eventually decreases such that $\rho_{eff}$ approaches zero in late times. This suggests that the dark energy density reduces to almost zero during later periods (See Fig. \ref{rho}). The energy density of scalar field $\rho_{\phi}$ increases with time but eventually collapses to $0$ at $z=-1$ (See Fig. \ref{rhophi}). The image in Fig. \ref{pressure} depicts that the highly negative isotropic pressure, $p_{eff}$, decreases with time. According to the standard cosmology, cosmic acceleration is associated with negative values of cosmic pressure.
Energy conditions (ECs) are viewed as a sound generalization to the entire EMT of motion, stating that the universe's energy density can never be negative. The use of ECs in examining the viability of numerous significant space-time singularity problems, such as those involving black holes and wormholes, is widespread. Numerous ECs are frequently employed in GR and whose viability can be examined using the well-known Raychaudhuri equation. The energy conditions can generally be expressed in one of two ways: (i) geometrically, where the ECs are expressed in terms of the Ricci tensor or the Weyl tensor, or (ii) physically, where the ECs are expressed in terms of the stress energy-momentum tensor or in terms of the energy density, which functions as the time-like component and pressures $p_i$, i = 1, 2, 3, which represent the three-space-like components. The typical point-wise ECs in GR are defined as (a) Null Energy Condition : NEC: $ \rho+p_i \geq 0, \forall i$; Strong Energy Condition : SEC: $ \rho+\sum_{i=1}^3 p_i \geq 0, \forall i $; Dominant Energy Condition : DEC:  $ \rho \geq 0, \rho \geq |p_i|, \forall i, $ where $i=$ 1,2,3; Weak Energy Condition :WEC:  $ \rho \geq 0 $, $ \rho+p_i \geq 0, \forall i $. 
 
\begin{figure}[ht]
 \subfloat[]{
	\begin{minipage}[c][1\width]{
	   0.3\textwidth}
	   \centering
	   \includegraphics[width=1.05\textwidth]{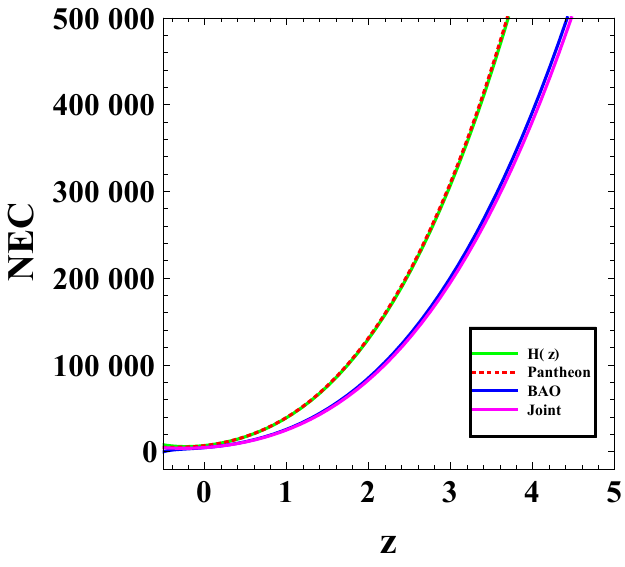}
    \label{nec}
	\end{minipage}}
 \hfill 	
  \subfloat[]{
	\begin{minipage}[c][1\width]{
	   0.3\textwidth}
	   \centering
	   \includegraphics[width=1.05\textwidth]{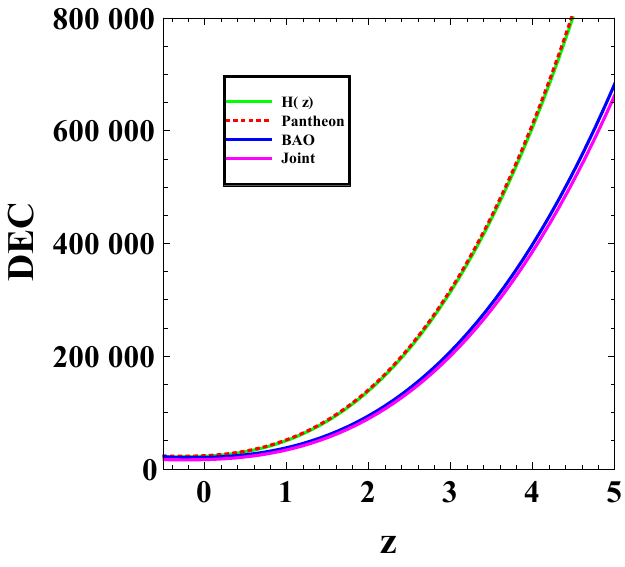}
    \label{dec}
	\end{minipage}}
 \hfill	
 \subfloat[] {
	\begin{minipage}[c][1\width]{
	   0.3\textwidth}
	   \centering
	   \includegraphics[width=1.05\textwidth]{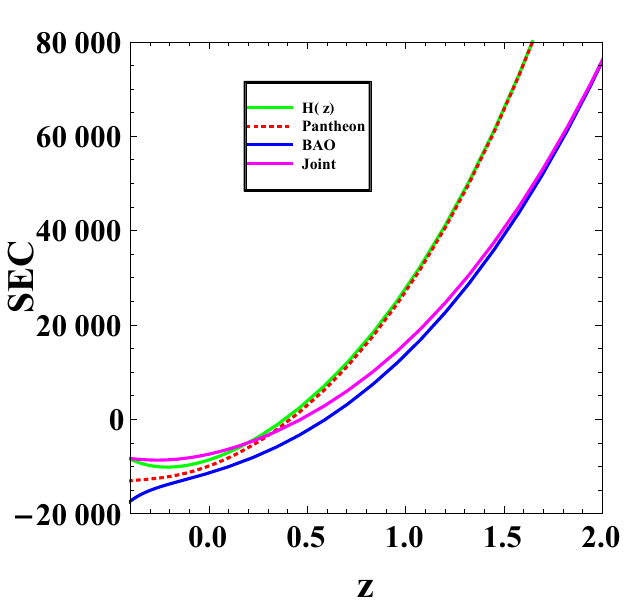}
    \label{sec}
	\end{minipage}}
\caption{ Graphical behaviour of NEC, DEC, and SEC respectively.}
\label{EC}
\end{figure}

 By examining Fig. \ref{EC}, it can be seen that our model demonstrates accelerating expansion in the universe as NEC and DEC are satisfied, but SEC is violated. The violation of SEC suggests that exotic matter may be present in the universe that produces anti-gravitational effects, which are responsible for the universe's accelerated expansion (See Fig. \ref{sec}).

$ Om(z) $ stands for Om diagnostic \cite{Sahni:2008xx,Zunckel:2008ti}. It differentiates between the basic $ \Lambda $CDM and several dark energy models. Om is easier to reconstruct from observational data since it just uses the first derivative of the scale factor via the Hubble parameter. $ Om(z) $ is defined as 
\begin{equation}\label{39}
Om(z)= \frac{(\frac{H(z)}{H_0})^2-1}{z(z^2+3z+3)}.
\end{equation}
This tool suggests three different types of dark energy behaviour: quintessence type $ ( \omega > -1) $ corresponding to its negative curvature (i.e. below the $ \Lambda CDM $ line), phantom type $ ( \omega < -1) $ relating to its positive curvature (i.e. above the $ \Lambda CDM $ line), and $ Om(z) = \Lambda CDM $ corresponding to zero curvature. Fig. \ref{om} reveals that the trajectories of $Om(z)$ increase as redshift $z$ decreases, indicating a negative curvature pattern that confirms our model as a quintessence DE model for all observational datasets. Our model collapses at 1 in the future, i.e., at $z=-1$.

\section{ Thermodynamical behaviour}\label{Sec5}
This section examines applying thermodynamics's generalized second law (GSL) to our model \cite{Sheykhi:2008qs, Setare:2008bb}. It is important to note that thermodynamics dictates that the entropy of isolated systems cannot decrease. We investigate the computation of the total entropy $S$ for this model, assuming that the apparent horizon is connected to temperature and entropy in a way that is similar to the black hole event horizon. Consequently, the GSL requires that the total entropy $S$ encompasses the entropy of all sources \cite{Barrow:1988yc}. As the Universe evolves, the rate of the entropy change for both the fluid within it and the horizon must remain at zero or increase. In our examination of recent and upcoming times, the total entropy is the combination of the entropy contribution from matter $(S_m)$ and the apparent horizon $(S_h)$, such that $S =S_m+ S_h$, where $S_h=\frac{k_B A}{4l_{pl}^2}$ represents the entropy of the apparent horizon and $S_m$ represents the entropy of pressure-less matter. The horizon's area and Planck's length are denoted by $A$ and $l_{Pl}$, respectively, and the Boltzmann constant is represented by $k_B$. The apparent horizon's size can be determined using the formula $A = 4\pi r_h^2$, where $r_h =\frac{1}{\sqrt(H^2+ka^{-2})} $. Since we're focusing on a spatially flat model $(k = 0)$, we can assume that $r_h = H^{-1}$. This means that the entropy of the horizon can be expressed as
\begin{equation}\label{31}
    S_h=\frac{k_B \pi}{l_{pl}^2H^2}.
\end{equation}
The Gibbs relation describes the entropy of matter within the dynamic apparent horizon as \cite{Li:2015bza, Izquierdo:2005ku}
\begin{equation}\label{32}
TdS_m=d(\rho V)+pDV= V d\rho+(p+\rho)dV.
\end{equation}
The equation assumes that the temperature of the fluid, denoted by $T$, is equal to the temperature at the horizon $(T_h)$, which is calculated as $T_h = \frac{1}{2\pi r_h}$, where $r_h$ is the radius of the horizon. $V$, on the other hand, represents the spatial volume bounded by the horizon and is equal to $\frac{4 \pi}{3}r_h^3 $.
Differentiating Eqs. (\ref{31}) and (\ref{32}) and adding them gives us the value of $\dot{S_h} + \dot{S_m}$.
\begin{figure}[ht]\centering
	\subfloat[]{{\label{a}\includegraphics[scale=0.36]{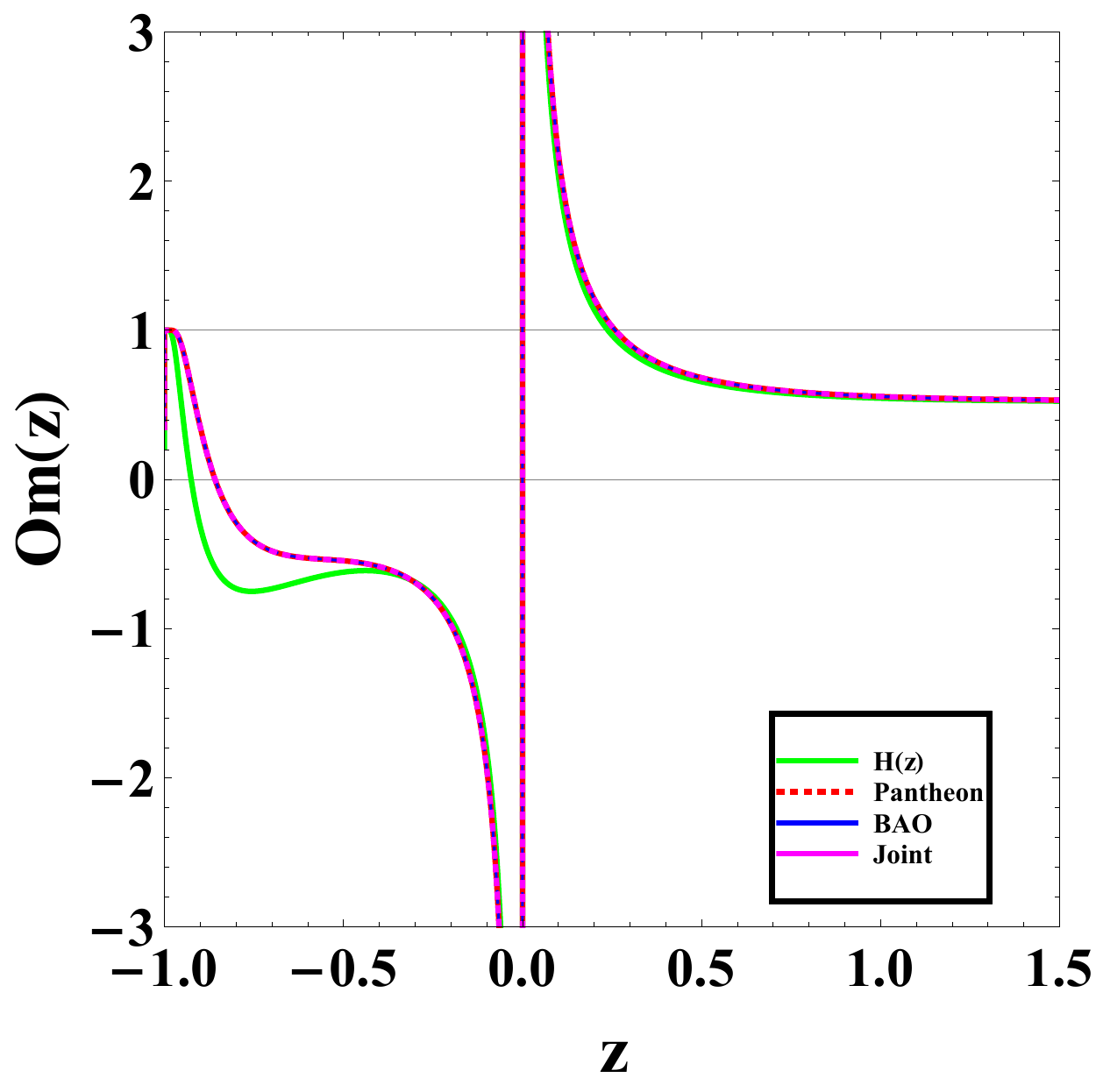}}
\label{om}} \hfill
	\subfloat[]{{\label{b}\includegraphics[scale=0.41]{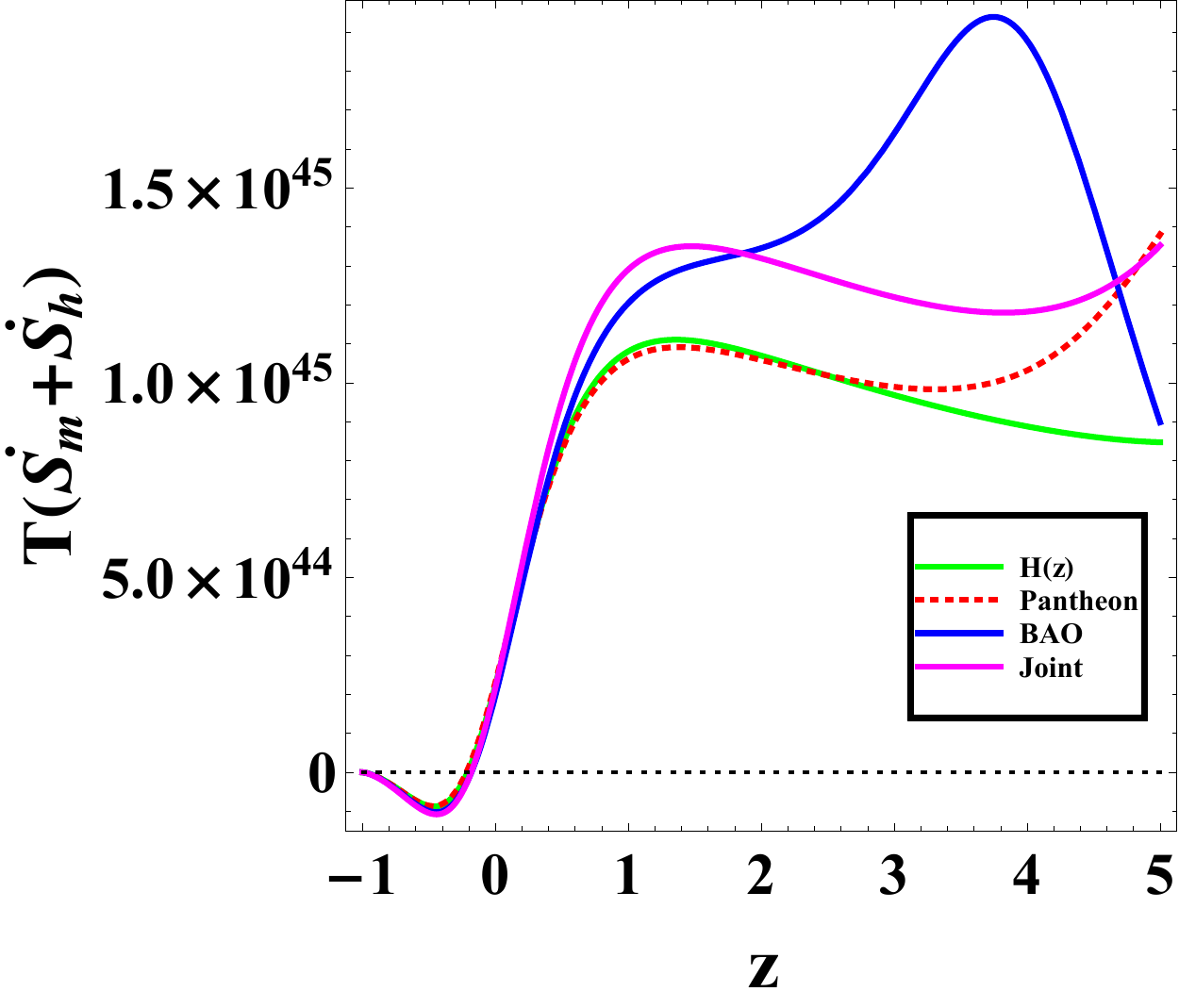}} 
 \label{Entropy}}
	\caption{ The plots depict the behaviour of the Om diagnostic and the variation of the temperature of the fluid in terms of the sum of the rates of change of entropy for matter and the apparent horizon \textit{i.e.} T($\dot{S_h} + \dot{S_m}$). }
 \label{om-entropy}
\end{figure}
 $\dot{S_h} + \dot{S_m}$ remains non-negative throughout the entire history of the universe. This implies that the sum of the rates of change of entropy for matter and the horizon denoted as $\dot{S_h}$ and $\dot{S_m}$ respectively, is always greater than or equal to zero, as shown in Fig. \ref{Entropy}. Therefore, the Generalized Second Law of thermodynamics is always satisfied for a spatially flat FRW universe containing pressure-less dark matter within the region enclosed by the apparent horizon.

\subsection{Limitations and Assumptions}\label{Sec5a}

While our obtained model is consistent with recent observational datasets, it does have certain limitations:
\begin{itemize}
  \item  our model expands with eternal acceleration from high redshift to low redshift (see Fig. \ref{DP}),
 \item  in Fig. \ref{jerk}, we see that the model approaches $\Lambda$CDM at present ($ z \to 0 $),
 \item  the EoS parameter $ \omega_{\phi} $ is consistent with the $\Lambda$CDM for high redshift $ z $ (see Fig. \ref{wphi}),
 \item in Fig. \ref{wtotal}, we see that our model shows dust-filled universe for the redshift range $ z>4.52 $.  Also, the other properties of the model can be seen in Table II in detail,
 \item  the energy densities $ \rho_{\phi} $, $\rho_{eff}$, and the isotropic pressure $p_{eff}$ approach zero as $ z \to-1 $,  which indicates that this model seems to collapse in the later times (see Fig. \ref{rho-p}),
 \item our model demonstrates accelerating expansion in the universe as NEC and DEC are satisfied, but SEC is violated. The violation of SEC suggests that some exotic matter may be present in the universe that produces anti-gravitational effects, which are responsible for the universe's accelerated expansion (See Fig. \ref{EC}).
\end{itemize}

\begin{table}[htbp]
\caption{ \textbf{  Observational analysis of the physical parameters of the model}}
\begin{center}
\label{tabparm2}
\begin{tabular}{l c c c r} 
\hline\hline
\\ 
 Observational datasets  &   Redshift range   &  ~~~  $ EoS~ parameter $~~~  &  Physical behaviour
\\
\hline 
\\
 $ H(z) $  &  ~~$ z > 4.52 $ ~~ &   $ \omega_{eff}=0 $   &  $ Dust~filled~ universe $

\\
  &   ~~$ 0.4968 > z > -0.6766 $ ~~ &  $ -\frac{1}{3}>\omega_{eff}>-1 $ &  $ Quintessence-like $
\\
 &   ~~$ z \approx -0.6766 $ ~~ &  $ \omega_{eff}=-1 $ &    $ \Lambda $CDM
\\
 & ~~$ -0.6766 > z >-1 $~~  &  $ \omega_{eff}<-1 $  &   $ Phantom-like $
\\
\hline 
\\
$ Pantheon $ &   ~~$ z > 4.52 $ ~~  &  $ \omega_{eff}=0 $ &   $ Dust~filled~universe $
\\
 &  ~~$ 0.4358 > z > -0.6462 $ ~~  & $ -\frac{1}{3}>\omega_{eff}>-1 $ &   $ Quintessence-like $
\\
 &   ~~$ z \approx  -0.6462 $ ~~  & $ \omega_{eff}=-1 $ &   $ \Lambda $CDM
\\
 &  ~~$ -0.6462 > z >-1 $~~   &  $ \omega_{eff}<-1 $  &   $  Phantom-like $
\\
\hline 
\\
 $ BAO $ &  ~~ ~~$ z > 4.52 $ ~~   &  $ \omega_{eff}=0 $  &   ~~~ $ Dust~filled~ universe $
\\
 $  $ &  ~~$ 0.6187 >z > -0.4633 $ ~~  &  $ -\frac{1}{3}>\omega_{eff}>-1 $ &   $ Quintessence-like $
\\
 &   ~~$ z \approx  -0.4633 $ ~~  & $ \omega_{eff}=-1 $ &   $ \Lambda $CDM
\\
 $  $ &  ~~$ -0.4633 > z > -1 $~~   & $ \omega_{eff}<-1 $  &   $  Phantom-like $
\\
\hline 
\\
 $ Joint ~data $ &   ~~$ z > 4.52 $ ~~&  $ \omega_{eff}=0 $  &  ~~~ $ Dust~filled~ universe $
\\
$  $ &  ~~$ 0.5120 > z > -0.6614 $ ~~ & $ -\frac{1}{3}>\omega_{eff}>-1 $ &   $ Quintessence-like $
\\
&   ~~$ z \approx -0.6614 $ ~~&  $ \omega_{eff}=-1 $  &  $ \Lambda $CDM
\\
$  $ &  ~~$ -0.6614 > z > -1 $~~   &  $ \omega_{eff}<-1 $   &   $  Phantom-like $
\\
\hline\hline 
\\
\end{tabular}    
\end{center}
\end{table}

\section{Conclusion}\label{Sec6}
 We have examined a cosmological model of a homogeneous and isotropic universe within the $ f(R, T) $ gravity framework using a specific differential equation, which is parameterized in terms of the energy density of the scalar field $\rho_{\phi}$. By employing this parameterization, we have reconstructed the Hubble parameter, $ H(z) $, to analyze the universe's evolution in a flat FLRW space-time. The four model parameters are constrained using the Hubble, Pantheon, BAO, and their joint data. Certain significant cosmological parameters have been formulated as a function of redshift $ z $ and illustrated through graphical representation to comprehend cosmic evolution. Our obtained model has also been compared with the standard model $\Lambda$CDM. 

 The contour plots depict the best-fit values of the model parameters $\Omega_{m0}$, $\Omega_{\phi 0}$, $H_0,$ and $ \alpha $ with $1-\sigma$ and $2-\sigma$ errors for the Hubble, Pantheon, BAO, and their joint datasets (see Figs. \ref{Hub}-\ref{Hz+Pan}). The MCMC analysis indicates that the optimal values of the model parameters derived from the Hubble dataset are slightly dissimilar to those obtained from the Pantheon dataset. The best-fit values obtained from these datasets align with the $\Lambda$CDM model. The error bar plots for the Hubble and SNeIa Union 2.1 compilation datasets show a good fit when our model is correlated with $\Lambda$CDM in Fig. \ref{H-mu}. Our model is ever-accelerating, and its present value $q_0 \approx -1.004$ aligns with the available observational data. The behaviour of the jerk parameter is consistent with the $\Lambda$CDM at present $z=0$ (see Fig. \ref{DP-J}).
 
 The evolution of the EoS parameter of the scalar field, $\omega_{\phi}$, seems to coincide with the $\Lambda$CDM in the past and the quintessence region at present before collapsing in the future for a short period and emerging to  $\omega_{\phi} \to -1$ at $z=-1$ (See Fig. \ref{wphi}). The behaviour of $\omega_{eff}$ is depicted in Fig. \ref{wtotal}, which can be observed that it remains in the quintessence region in the past and at present before collapsing to $0$ at $ z=-1 $. Fig. \ref{rho-p} depicts the behaviour of energy density, the energy density of the scalar field, and pressure. $\rho_{eff}$ and $\rho_{\phi}$ decrease with time and collapses to $0$ in future i.e at $z=-1$. The effective pressure, $p_{eff}$, is negative at present ($z=0$) and attains an extremum value before collapsing to $0$ in the late times ($ z=-1 $). Therefore, this model represents the quintessence model at present. The physical interpretation of the model's behaviour in the various redshift ranges and its convergence to dust-filled universe in the infinite future has been discussed in Table \ref{tabparm2}.
 
The null, dominant, and strong-energy conditions have been determined through this analysis. The SEC is critical in these conditions as it characterizes gravity's attractive or repulsive behaviour in the context of modified gravity theory. Based on the observed values, it has been seen that NEC and DEC conform to the conditions derived from the Raychaudhuri equation, while SEC does not meet these conditions. The SEC violation indicates the universe's accelerated expansion (See Fig. \ref{EC}). Om diagnostic tool compares our obtained model with the standard model $\Lambda$CDM. The model's behaviour is depicted in Fig. \ref{om}, comparing the standard $\Lambda$CDM model and the SCDM model. This model shows a quintessence model in late times because the variation of the Om diagnostic curve has a negative slope in Fig. \ref{om}.

 Also, we have examined the $ GSL $ of thermodynamics. The entropy shows a positive value in the early stages of the evolution and a negative value in the later stages before converging to zero. Therefore, the $ GSL$  holds good during the initial stages of the evolution. Still, it becomes invalid during the later stages (See Fig. \ref{Entropy}). Moreover, Our model exhibits quintessence-like behaviour in the redshift range $ 0.5120 > z > -0.6614 $, and Phantom-like behaviour in the redshift range $ -0.6614 > z > -1 $ according to the joint data observations. Thus, our model is highly unstable before converging to the dust-filled universe and intersecting the quintom line ($\Lambda$CDM) in the infinite future ($ z=-1 $). Finally, our model represents an eternal accelerated expanding model up to the redshift $ z \approx -0.6614 $ ($ \Lambda $CDM).

\vskip0.3in

\textbf{\noindent Acknowledgements}\\
 J. K. Singh wishes to thank Dr. Saddam Hussain, Department of Physics, Indian Institute of Technology, Kanpur, Uttar Pradesh 208016, India, for fruitful discussions.

\end{document}